\documentclass[10pt, letterpaper, onecolumn]{IEEEtran}
\usepackage{graphics}
\usepackage{amsmath}
\usepackage{amssymb}
\usepackage{graphicx}
\usepackage{cite}
\usepackage{epstopdf}
\usepackage{datetime}

\newtheorem{theorem}{\bf Theorem}

\newtheorem{definition}{\bf Definition}

\DeclareMathAlphabet{\mathssf}{OT1}{cmss}{m}{sl}

\newcommand{\lo}[1]{\log_2\left(#1\right)}
\newcommand{\lon}[1]{\ln\left(#1\right)}

\newcommand{\rxpower}{P_R}
\newcommand{\txpower}{P_T}
\newcommand{\pe}{{\langle P_e\rangle}}
\newcommand{\pei}{{\langle P_{e,i}\rangle}}

\title{Towards a communication-theoretic understanding\\\vspace{-0.25in} of system-level power consumption\thanks{Results in this work have been presented in part at ISIT 2008~\cite{greencodes} and ISTC 2010~\cite{ISTC10Paper}. This work is the outcome of many discussions with students and faculty at the Berkeley Wireless Research Center and the Wireless Foundations. Those with Elad Alon, Bora Nikolic, Hari Palaiyanur, Jan Rabaey and Matt Wiener are especially acknowledged. We are also thankful for comments on the paper draft by Sudeep Kamath, Sameer Pawar and Salim El Rouayheb. This work is supported by NSF grants CCF-0917212 and CNS-0403427.}}
\author{\vspace{-0.15in}Pulkit Grover, Kristen Ann Woyach and Anant Sahai\\\vspace{-0.1in}University of California, Berkeley, Berkeley, CA-94720}

\begin{document}\maketitle
\vspace{-0.8in}

\begin{abstract}
Traditional communication theory focuses on minimizing transmit power. However, communication links are increasingly operating at shorter ranges where transmit power can be significantly smaller than the power consumed in decoding. This paper models the required decoding power and investigates the minimization of total system power from two complementary perspectives.

First, an isolated point-to-point link is considered. Using new lower bounds on the complexity of message-passing decoding, lower bounds are derived on decoding power. These bounds show that 1) there is a fundamental tradeoff between transmit and decoding power; 2) unlike the implications
of the traditional ``waterfall'' curve which focuses on transmit power, the \textit{total} power must diverge to infinity as error probability goes to zero; 3) Regular LDPCs, and not their known capacity-achieving irregular counterparts, can be shown to be power order optimal in some cases; and 4) the optimizing
transmit power is bounded away from the Shannon limit.

Second, we consider a collection of links. When systems both generate and face interference, coding allows a system to support a higher density of transmitter-receiver pairs (assuming interference is treated as noise). However, at low densities, uncoded transmission may be more power-efficient in some cases.
\end{abstract}

\IEEEpeerreviewmaketitle

\vspace{-0.3in}

\section{Introduction}

\vspace{-0.1in}

The first transistor started the field of modern circuits~\cite{Transistor} around the same time as the foundations of modern communication theory were being laid~\cite{ShannonOriginalPaper}. These twin discoveries led the world into the digital revolution we witness today. Traditionally, the development of wireless communication has followed a division of labor: the design of techniques (e.g. error-correcting codes) that minimize transmit power was complemented by the development of power and area-efficient circuit-infrastructure that could process signals at the transmitter and the receiver. This division of labor was justified at the time: since the distances of communication were large for most practical applications (e.g. deep-space communication~\cite{Massey92DeepSpace} was very influential in the development of the theory), transmit power dominated the processing power consumed in circuits, and therefore received most of the theoretical attention. 

However, two developments are upsetting the justification for this division of labor. The first is the development of capacity-approaching sparse-graph codes with low decoding complexity. Because the decoding algorithms for sparse-graph codes have an efficient and intuitive parallel implementation, circuit engineers can design codes and decoding architectures simultaneously (e.g. ``architecture-aware'' LDPC codes in~\cite{MansourShanbhag}) so that the decoders are easy to implement and still promise good performance.

\begin{figure}[htb]
\begin{center}
\includegraphics[width=3.6in]{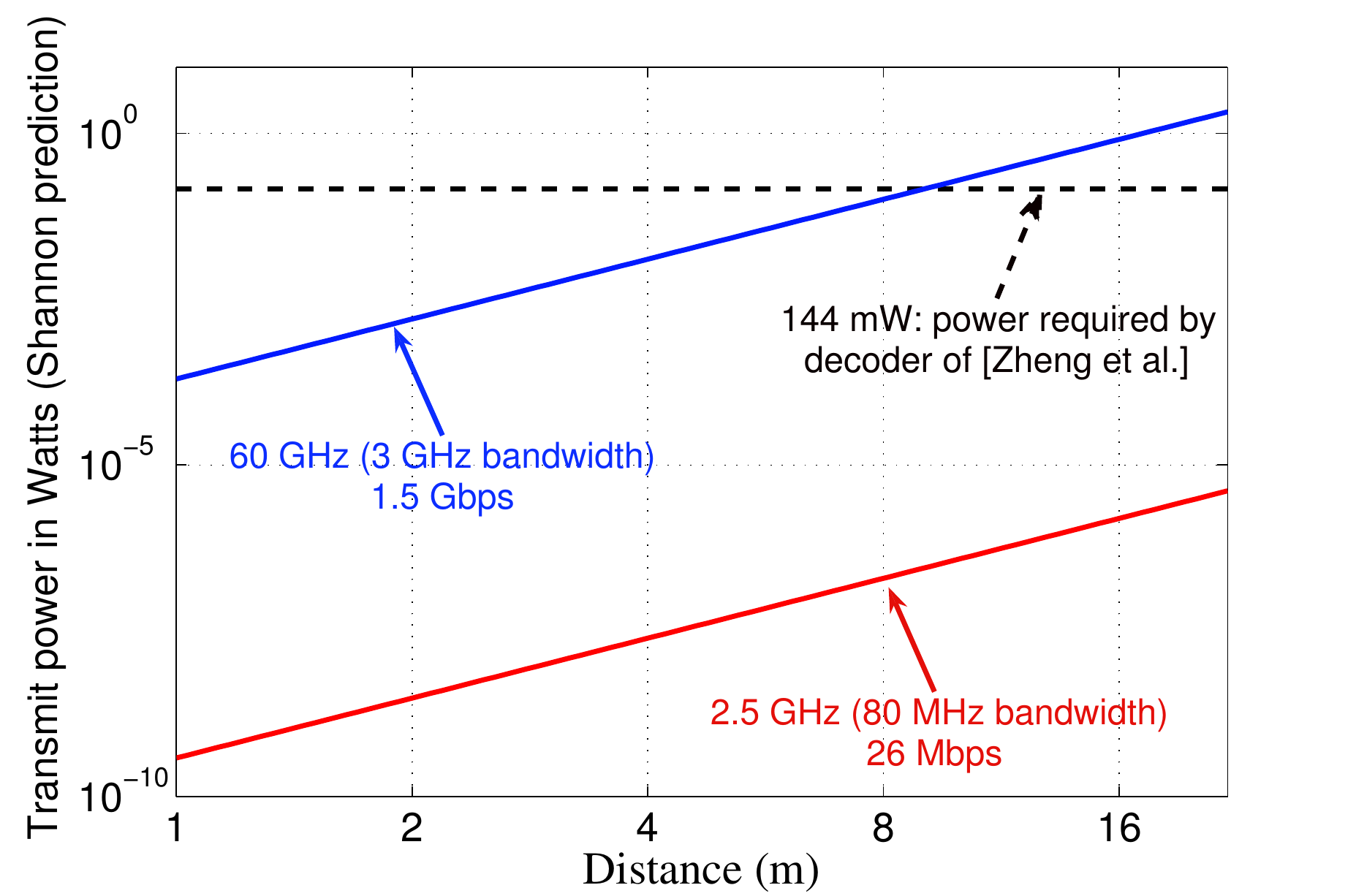}
\caption{The required transmit power for two short-distance bands of interest. The ISM band, centered at $2.5$ GHz shows the required power for bluetooth applications ($80$ MHz bandwidth) for a data-rate of $26$ Mbps. The $60$ GHz band presents the upcoming high-bandwidth high-throughput wireless paradigm. The bandwidth is large ($3$ GHz), and the throughput is $1.5$ Gbps. Path-loss exponents are assumed to be $3$ (indoor environment), and the noise figure is $3$ dB. Most applications today lie somewhere between the two curves. Observe that even for $1.5$ Gbps link, the transmit power is not more than a few hundred milliwatts for a distance of $3$ m. Many of these applications are designed for even smaller distances, where the transmit power is only a few tens of milliwatts.}
\label{fig:transmitpower}
\end{center}
\end{figure}

\begin{figure}[htbp] 
   \centering
   \includegraphics[width=2.5in]{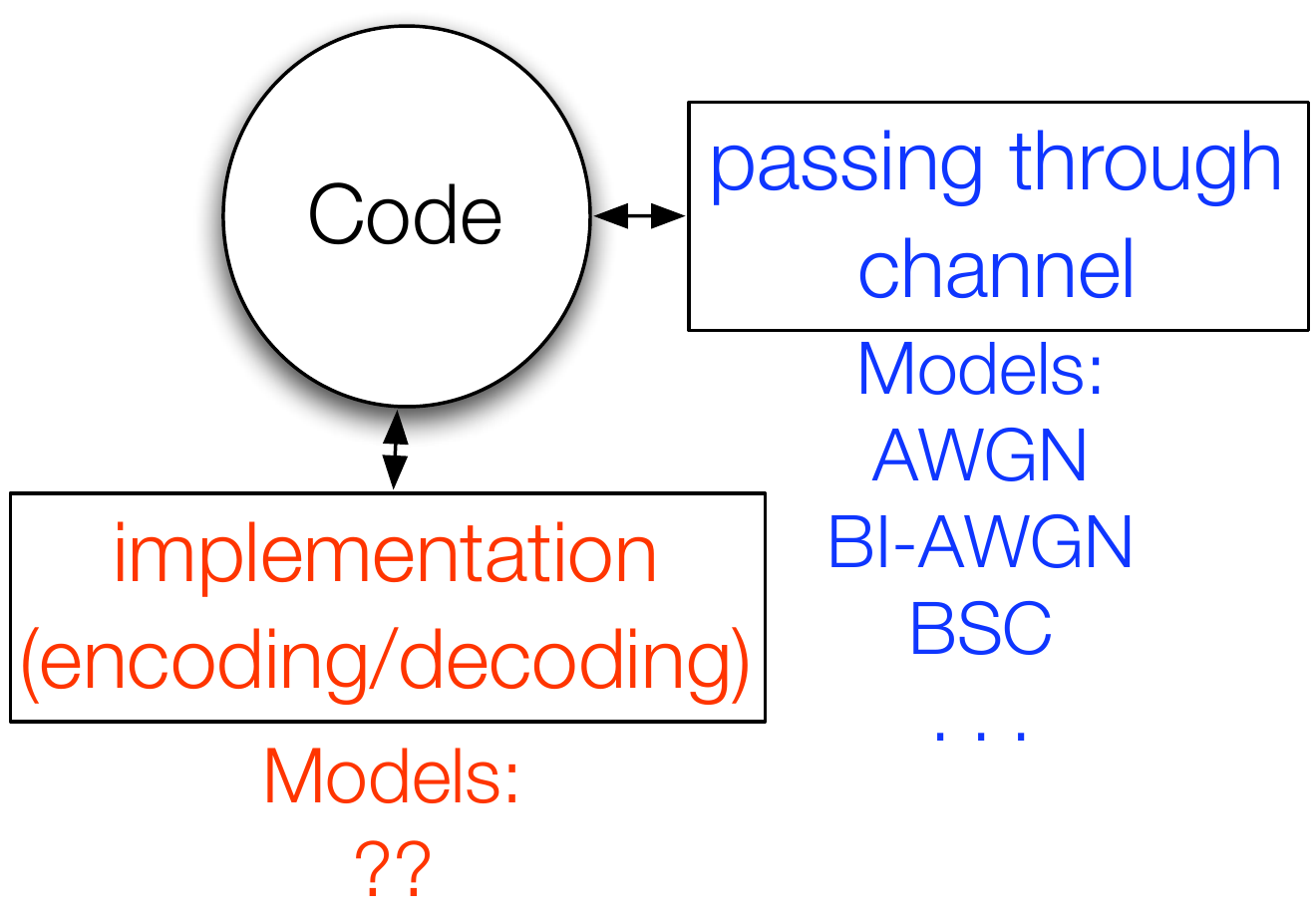} 
   \caption{A code is an abstract mathematical object that interfaces with the physical world as the codewords pass through a channel, and at the encoding/decoding implementations. While channel models are well studied, models of decoding implementations are not commonly investigated. Power is consumed at both of these interfaces.}
   \label{fig:interfaces}
\end{figure}

Perhaps more significant is the second development: battery-powered devices that communicate at short distances (e.g. bluetooth, wireless sync, personal area networks, etc.). As Fig.~\ref{fig:transmitpower} illustrates, for distances smaller than $10$ meters, transmit power is often comparable to, or much smaller than, the power required by most state-of-the-art decoders (see example implementations in~\cite{zhengyaJournal,zhengyaThesis}). Indeed, uncoded transmission is commonly used (e.g. in Wireless LANs~\cite{hanzo}, 60 GHz band~\cite{marcu}, etc.) to reduce power consumed in processing despite increased transmit power. As shown in Fig.~\ref{fig:interfaces}, a code interfaces to the physical world not only in the channel, but also in the encoding/decoding implementation. Both of these interfaces consume power. \textit{Just as we have channel models that help us understand transmit power, we need models of decoding implementation in order to understand decoding power}. A corresponding \textit{total} power extension of communication theory that unifies transmit and processing power is required to guide such implementation efforts. Without such a theory, We would not even know, for example, whether approaching traditional Shannon-capacity is still a worthy goal to pursue. 

The difficulty in developing a unified theory lies in developing good models for power consumed in processing. How do we abstract the power consumed by various possible processing algorithms, circuit designs, and architectures? As a first step, the authors in~\cite{goldsmithbahai,massaadJournal} model the transmitter and receivers as black-boxes that consume a fixed amount of energy per unit time powered `on'. Whether the problem is one of constellation design, as considered by Cui, Goldsmith and Bahai~\cite{goldsmithbahai}, or of coded transmissions, as considered by Massaad, Medard and Zheng~\cite{massaadJournal}, the message is the same: since keeping systems powered `on' consumes processing energy, transmissions should be ``bursty'' --- both receiver and transmitter are shut `off' for some time --- in order to reduce processing power. However, the power required in transmission increases exponentially with burstiness (because capacity scales logarithmically in power), so the transmission must not be too bursty. The existence of an optimal non-zero level of burstiness is surprising: traditional transmit power analysis~\cite{VerduCost} for a non-fading channel\footnote{For fading channels, traditional analysis suggests that bursty transmissions can help~\cite{BurstyHajek}.} predicts that the transmission rate should be made as small as possible, and the signals least bursty, for minimum energy consumption.

Because it lumps together all of the power used in processing the signal, the black-box model does not yield much insight into code choice or decoder design. It has been observed~\cite{marcu} that for high data rate communication, the  power required for decoding tends to dominate the other sinks of power (e.g. ADC, DAC, encoding, modulation/demodulation, amplification, etc.) in processing at the transmitter and the receiver\footnote{The fact that decoding power is the dominant sink of processing power is what allows uncoded transmission to significantly reduce system power consumption in the settings of interest in~\cite{marcu}.}. As a first-order approximation, the theory of system-power minimization can therefore focus on just decoding power. Still, the existence of many different codes and multiple decoding algorithms for each code complicates the modeling problem. Modern coding theory ensures that this complication will only grow as it continues to be enriched by increasingly practical codes (e.g. turbo codes, LDPC codes, IRA codes, ARA codes, etc.~\cite{ModernCodingTheory}) that approach capacity, have low-decoding complexity, and have been implemented with various decoding architectures (see~\cite{zhengyaThesis} for some of the possible architectures).

One approach to deal with the plethora of codes and decoding architectures is to perform an empirical study of existing codes and decoders. To the best of our knowledge, the work of Howard \textit{et al}~\cite{HowardSchlegel} is the first to attempt a comprehensive survey of coding/decoding strategies from a \textit{total} (transmit and decoding) power\footnote{Again, the attention is limited to transmit and decoding power because they typically dominate other power sinks.} perspective. They use empirical power-consumption numbers for certain chosen code/decoder implementations at moderately low probabilities of error. They observe that at sufficiently small distances (depending on the choice of the code and the decoding), the increase in power consumption due to decoding is larger than the savings in transmit power because of coding. This provides a justification for the use of uncoded transmission in cases such as~\cite{marcu}. 

This empirical approach breaks down when the application at hand desires a different error probability, or operates in an environment with different path loss. Do the same codes continue to be the most power-efficient? Howard \textit{et al}~\cite{HowardSchlegel} chart out the performance of a few families of codes at different error probabilities. However, even if all existing possibilities could be listed for the designer, empirical studies cannot rule out the possibility of better codes yet to be discovered. Furthermore, short-distance communication need not happen in isolation. Other communication links in the same frequency band will also complicate matters. Just as without Shannon-theory, empirical approaches would have been insufficient on their own for designing power-efficient long-distance communication systems, a  theoretical framework is required to guide the code/decoder design to minimize total power consumption in the short-range context. 

In this paper we take the first steps towards such a theoretical foundation. We examine the problem from two perspectives: the simplest case of an isolated point-to-point link, and a collection of non-cooperating links transmitting simultaneously. Our model for the decoding process is based on the observation that practical decoders for modern codes are all extremely parallelized: they are all based on some form of \textit{message-passing} decoding (for instance, in belief-propagation~\cite{urbankecapacity}, likelihood values are passed as messages). Message-passing architectures have been abstracted in the VLSI-theory literature~\cite{lengauer} (starting with the pioneering work of Thompson~\cite{thompson}) by a model that closely resembles message-passing decoding. As shown in Fig.~\ref{fig:VLSIModel}, the architecture has Processing Elements (PEs) that perform the desired computation by passing messages to each other to access information computed by other PEs and/or stored in the register of another PE. 

Adapting this model to parallelized message-passing decoding, in a companion paper~\cite{ComplexityITPaper} we derive information-theoretic lower bounds on the \textit{neighborhood size} of a bit for decoding any code to attain a specified error probability while operating at a given gap from capacity. The basic idea is simple: the ``visible universe'' for decoding any bit in message-passing decoding is the set of nodes it could have communicated with directly or indirectly, \textit{i.e.}~its decoding neighborhood. ``Sphere-packing'' bounds in traditional information theory~\cite{BlahutHypothesis,SahaiDelayBlocklength} are lower bounds on error-probability given the entire block. However, if the visible universe for a bit is smaller, the error probability should decrease with the size of the visible universe, and not the entire block. The ``local'' sphere-packing derivation  in~\cite{ComplexityITPaper} formalizes this idea.

In Section~\ref{sec:isolated}, we focus on the case of an isolated point-to-point link. In Section~\ref{sec:vlsi}, we adapt Thompson's model to a model of decoder power consumption. In Section~\ref{sec:lowbound}, by making use of the connectivity constraint in the VLSI model of decoding, we translate our bounds on required neighborhood size in~\cite{ComplexityITPaper} to bounds on the required number of iterations. Although bounds on iterations have been derived by Sason and Wiechman~\cite{sason},  their bounds hold only for specific code families and specific decoding algorithm (belief-propagation). Thus conceptually, the bounds of~\cite{sason} fall between our completely code-and-decoding-algorithm-agnostic results and the completely empirical results of Howard \textit{et al}~\cite{HowardSchlegel}.

\begin{figure}[htbp] 
\begin{center}
   \includegraphics[width=3.6in]{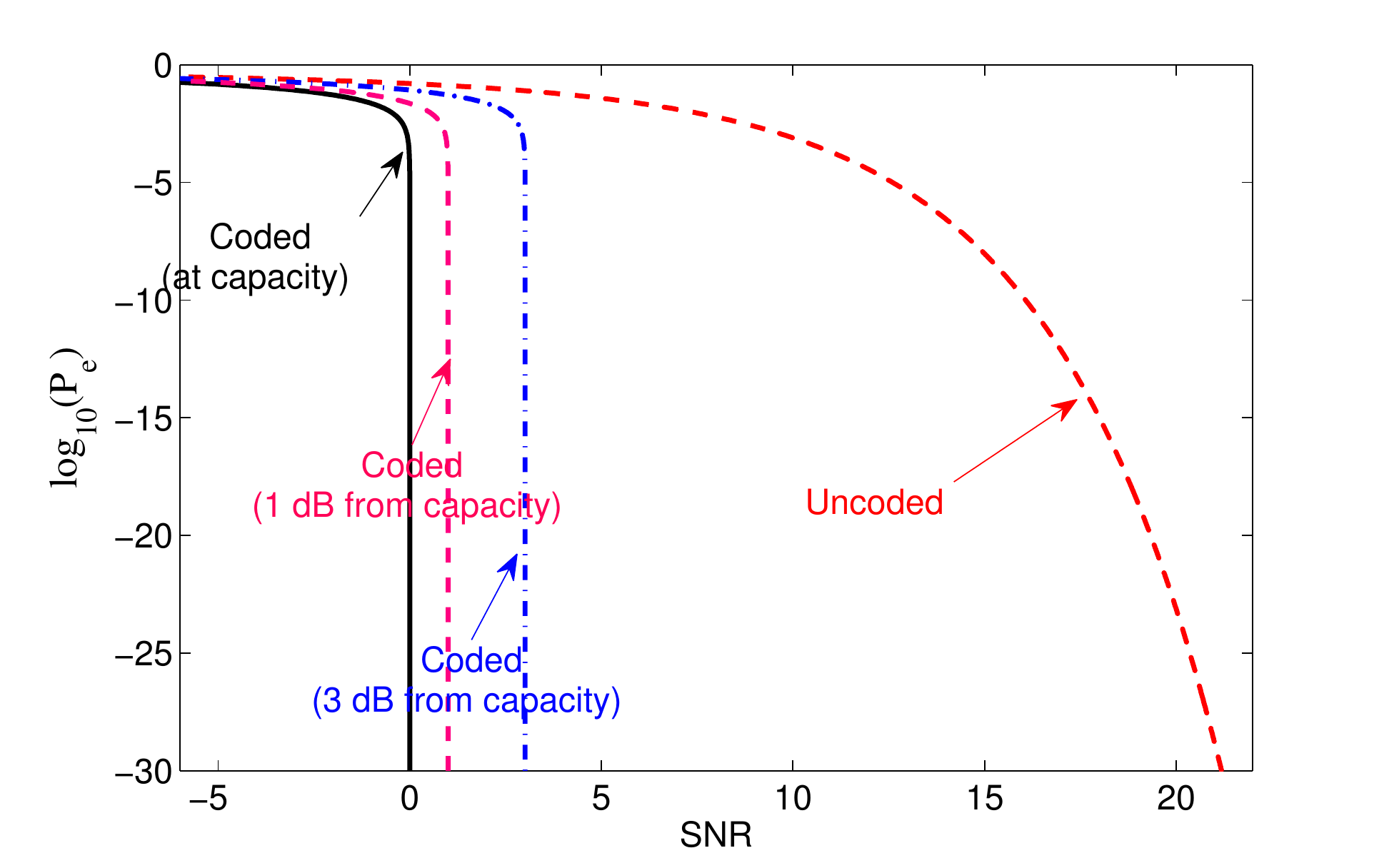} 
   \caption{The Shannon waterfall curve, which provides the minimum required SNR for small bit-error probabilities, predicts a bounded transmit power even as error probability converges to zero. In contrast, uncoded transmission requires that total power diverge to infinity. Also shown are required SNRs for codes that operate $1$ dB and $3$ dB away from capacity.}
   \label{fig:waterfall1}
   \end{center}
\end{figure}

Using our model of power consumption from Section~\ref{sec:vlsi} and the bounds on the number of iterations from Section~\ref{sec:lowbound}, we obtain lower bounds on decoding power consumption and total power consumption. Using these bounds, we show that the promise of Shannon-waterfall curves (see Fig.~\ref{fig:waterfall1}) ---  that the transmit power can remain bounded even as the error probability converges to zero --- does not extend to total power consumption. In contrast, our ``waterslide'' curves show that the total power must diverge to infinity as the error probability falls to zero. Further, there is a tradeoff between transmit power and decoding power, and the transmit power must therefore be strictly larger than the Shannon limit in order to keep the total power consumption small.

How good are these power bounds? In Section~\ref{sec:ldpc}, we show that \textit{regular} LDPCs (and \textit{not} their capacity-achieving counterparts\footnote{We observe that we do not claim that \textit{all} irregular LDPC codes are not order-optimal in this total power sense. As we will see, it is \textit{capacity-approaching} LDPC codes  (which have an extremely suboptimal decay in error-probability with number of iterations) that are not order-optimal.}) attain within a constant factor of the optimal total power even as $\pe\rightarrow 0$. Even so, the gap between the power they consume and our lower bounds is large enough to warrant a deeper look into the code design. Since regular codes cannot achieve capacity (under belief-propagation decoding)~\cite{burshteinBP}, and hence consume relatively large transmit power, redesigning of codes is required for short distance applications.

While point-to-point communication is a good starting point, in practical situations, short-distance links often exist in the company of other such links (for instance, the devices in license-free ISM bands are known to cause significant interference to each other~\cite{ISMInterference}). In the isolated link case, there are two ways of reducing the error-probability: increase the number of iterations at the decoder (thereby increasing the required decoding power), or simply increasing the transmit power and improving the SNR. What happens when a collection of point-to-point links communicate simultaneously in the same geographic area? An increase in transmit power is no longer sufficient because the transmit power of the interfering transmitters presumably increases as well, saturating the signal to interference and noise ratio (SINR). How can we increase SINR in this case? One way is to separate the transmitters by larger distances so that the collective interference is reduced. This is what MAC-protocols do, but it comes at the cost of a reduced \textit{density} of point-to-point links. Is there a better way, and if so, how well can we do? To investigate this, we introduce a simplistic model in Section~\ref{sec:collectionmodel}. Assuming that the interference is treated as Gaussian noise, we observe that coding has an important benefit in this case: it reduces the required transmit power thereby allowing the transmitters to operate closer to each other, increasing the supportable density of transmitter-receiver pairs. Bringing decoding power into the picture, in Section~\ref{sec:largepower} and~\ref{sec:finite} we propose an approach to investigate which code/decoder should be used, and whether we should use any coding at all. Within this context, the importance of the twin goals of coding theory shows up naturally: a code's \textit{gap from capacity} limits the maximum link density that can be supported, while high \textit{decoding complexity} (and consequently high decoding power) prevents the code from supporting good densities at small total power. Furthermore, even the coarse bounds explored here show that when the target link density is not maximal, it is best to operate codes away from capacity to minimize total power consumption.




\section{An isolated point-to-point link}
\label{sec:isolated}
\subsection{Models, definitions and problem formulation for an isolated point-to-point link}
\label{sec:vlsi}
\subsubsection{A VLSI model for decoding}

\begin{figure}[htb]
\begin{center}
\includegraphics[width=3.6in]{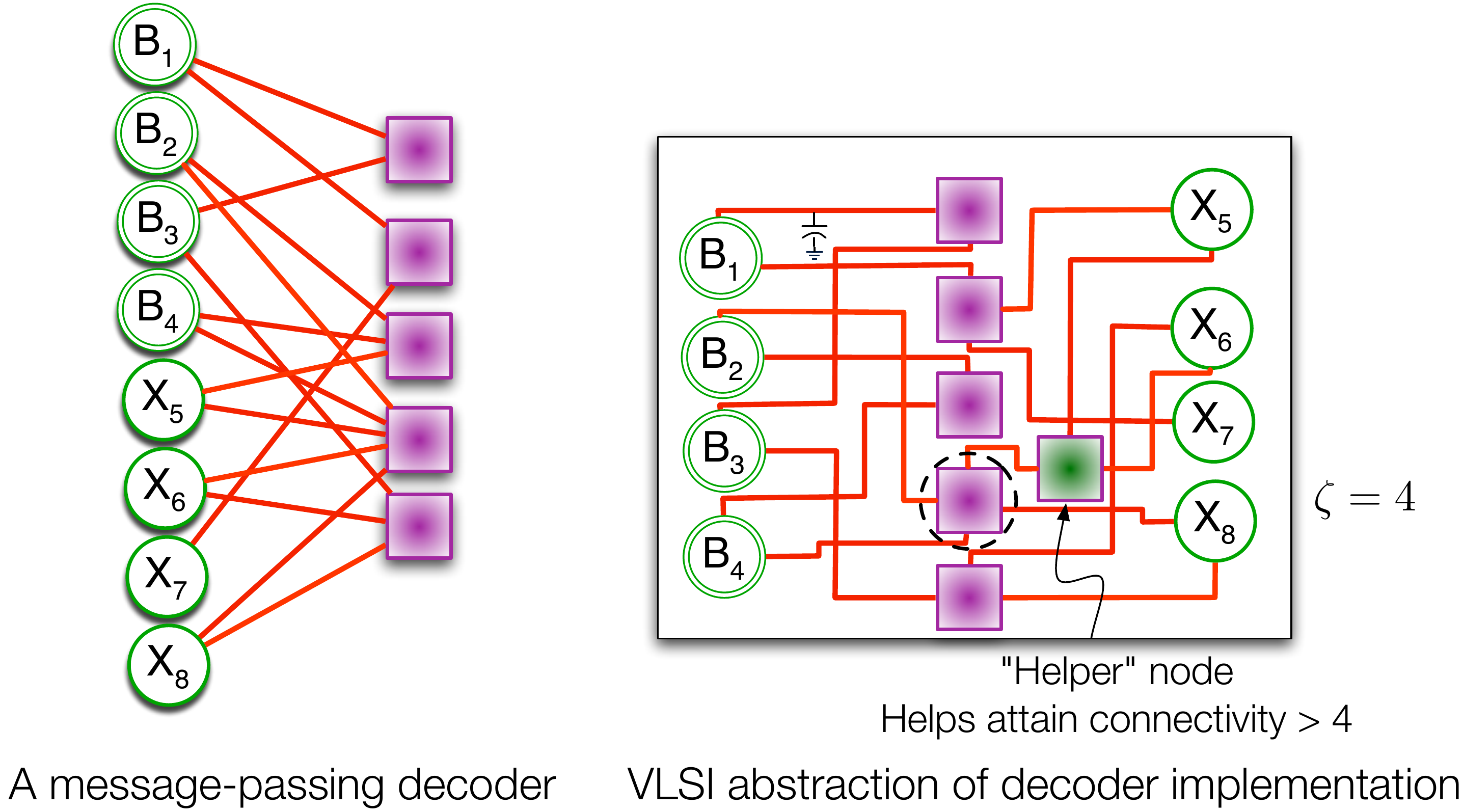}
\caption{An example VLSI model of a decoder. The Processing Elements (PEs) can be connected to at most $\zeta=4$ other PEs in this example. The same PE can act as any combination of output/input/helper PEs. The graph that models this chip is the obvious one --- each PE is represented by a node, and each wire by an edge.}
\label{fig:VLSIModel}
\end{center}
\end{figure}

A model for synchronous VLSI, called the ``VLSI model of computation,'' was introduced in~\cite{thompson,thompsonthesis} by Thompson. An example is shown in Fig.~\ref{fig:VLSIModel}. The model contains \textit{registers}, each of which has an accompanying \textit{processor} that can perform read/write operations on the register and other computational tasks. A register-processor pair is called a \textit{Processing Element} (PE)~\cite{thompson,lengauer}. The PEs are connected to each other by a set of \textit{wires}. At each \textit{iteration}\footnote{We use the term ``iteration'' instead of ``time-unit'' as used in~\cite{thompson,lengauer} to be consistent with the message-passing decoding literature. } --- a clock-cycle or a multiple thereof --- each PE communicates (sends and/or receives messages) with the PEs it is connected to. Thompson's goal was to be able to compare various architectures and algorithms against the best possible. 

%
This model has been used to explore the implementation and time complexity of the Discrete Fourier Transform~\cite{thompson}, sorting~\cite{cole88}, multiplication~\cite{sinhamultiply} and computing boolean functions~\cite{kramerboolean}. Universal VLSI models that can simulate any other VLSI implementation at some additional area and time cost have also been studied~\cite{BhattUniversal}. 

We adapt the VLSI model of computation to the problem of decoding an error correcting code. The PEs are either `message' PEs  that store the decoded bits after decoding,  `channel output' PEs that store the channel outputs, `helper' PEs that act as intermediaries of processing by improving connectivity (see Fig.~\ref{fig:VLSIModel}), or any combination thereof. We further assume that each PE is connected to at most $\zeta$ other PEs, an implementation constraint that arises from practical limitations on wire-density in microchips. 

To obtain lower bounds on power consumption, we first provide lower bounds on the time-complexity (\textit{i.e.} the number of iterations) by assuming a completely parallel decoding architecture. In practice, the required chip-area is often reduced by making the same PE act as two or more different PEs (of the same kind) in alternating iterations~\cite{zhengyaJournal}. In this paper, for simplicity, we will ignore this possibility and pretend that a fully parallel implementation is used. 

As was observed by Thompson~\cite{thompson}, one can abstract the decoder implementation as a decoder-connectivity graph. Each node in the graph represents a PE, and each wire connecting two PEs is an edge connecting the two nodes that represent the respective PEs. The structure of the resulting decoding graph imposes limitations on the information that can be passed between the PEs\footnote{The number of iterations --- a metric of complexity --- limits the information available to PEs. In this sense, the number of iterations is a measure of \textit{communication complexity}. Interestingly, Yao's seminal work on communication complexity~\cite{Yao} was indeed inspired by Thompson's model~\cite[Pg. 78]{UllmanBook}, as is also evidenced by another work of Yao~\cite{Yao2} which deals with bandwidth limited communication between PEs. Finding the decoding complexity can thus be viewed as finding the required communication complexity to decode message bits. It is this limitation that we exploit using a ``sphere-packing'' technique~\cite{BlahutHypothesis} in a companion paper~\cite{ComplexityITPaper} to obtain a lower bound on the decoding neighborhood size. }. We observe that this decoding overlay graph may not be the same as the constraint graph that conventionally~\cite{ModernCodingTheory} defines a sparse-graph code. In fact, parallelized graphical decoding algorithms have been developed for many codes not based on sparse-graphs, for instance Reed-Muller codes and polar codes~\cite{arikanBP}. To maintain greatest generality, we make \textit{no assumptions} on the code structure. 

\subsubsection{VLSI model of decoding power consumption}
For simplicity, we assume that each PE consumes a fixed $E_{node}$ joules of energy per iteration, irrespective\footnote{A more realistic model would also approximate the increase in $E_{node}$ with $R_{dec}$, but this is a subject of further investigation.} of the decoding throughput $R_{dec}$. We also assume that the $R_{dec}$ is the same as the data-rate $R_{data}$ across the channel\footnote{This is required to avoid buffer overflows at the decoder.}, measured in information-bits \textit{per second}. The data-rate in bits \textit{per channel-use} is denoted by $R_{ch}$.

The power received at distance $x$ meters is given by

\vspace{-0.35in}

\begin{equation}
\label{eq:rxpower}
P_R(\txpower,x)=\min\left\{\txpower,\frac{\txpower\lambda^\alpha}{x^\alpha}\right\},
\end{equation}
where $\lambda=\frac{c}{f_c}$ is the wavelength of transmission, $c=3\times 10^8$ meters per second is the speed of light, $f_c$ is the center frequency in Hertz, and $\alpha$ is the path-loss exponent, which is larger than $2$ in practical situations\footnote{This also rules out the unrealistic possibility of infinite interference at finite transmit powers, which is a mathematical consequence of using $\alpha=2$ in large networks.}. As a reality-check, we limit the maximum received power $P_R$ by the transmit power $P_T$. 
Let $\txpower=\xi_T \rxpower$ be the actual power used in transmission, where $\rxpower$ denotes the received power, and $\xi_T=\max\left\{1,\frac{x^\alpha}{\lambda^\alpha}\right\}$ represents the path-loss between the transmitter and the receiver. Let let $P_D$ be the power  consumed in the operation of the decoder. In this paper, we ignore the power consumed in encoding in the hope that it is much smaller than the decoding power. In the spirit of~\cite{Vasudevan}, we assume that the goal of the system designer is to minimize a weighted combination $P_{total} =  \xi_T \rxpower+ \xi_D P_D$ where the vector $\vec{\xi}=(\xi_T,\xi_D)$ has strictly positive elements. The weights can be different depending on the application. $\xi_T$ is tied to the distance between the transmitter and receiver as well as the propagation environment. To understand why we include the weight $\xi_D$, consider the example of an RFID application. The energy used by the tag is also supplied wirelessly by the reader. If the tag is the decoder, then it is natural to make $\xi_D$ even larger than $\xi_T$ in order to account for the inefficiency of the power transfer from the reader to the tag. One-to-many transmission of multicast data is another example of an application that can increase $\xi_D$. The $\xi_D$ in that case should be increased in proportion to the number of receivers that are listening to the message. 

For any rate $R_{data}$ and average probability of bit-error $\pe > 0$, we
assume that the system designer will minimize the weighted combination above to get an optimized $P_{total}(\vec{\xi}, \pe, R_{data})$ as well as constituent $\txpower(\vec{\xi}, \pe, R_{data})$ and $P_D(\vec{\xi}, \pe, R_{data})$.


\subsubsection{Definitions and Notation}
\label{sec:definitions}
We use the conventional information-theoretic model (see
e.g.~\cite{Gallager}) of fixed-rate discrete-time communication with $k$ total information bits, $m$ channel uses, and the rate of $R_{ch} = \frac{k}{m}$ bits per channel use. As is traditional, the rate $R_{ch}$ (bits/channel-use) is held constant while $k$ and $m$ are allowed to become asymptotically large. $\pei$ is the average probability of bit-error of the $i$-th message bit and $\pe = \frac{1}{k} \sum_i \pei$ is used to denote the overall average probability of bit-error. No restrictions are assumed on the codebook aside from the obvious requirements imposed by the channel-input alphabet.

\begin{figure}[htb]
\begin{center}
\includegraphics[width=3.6in]{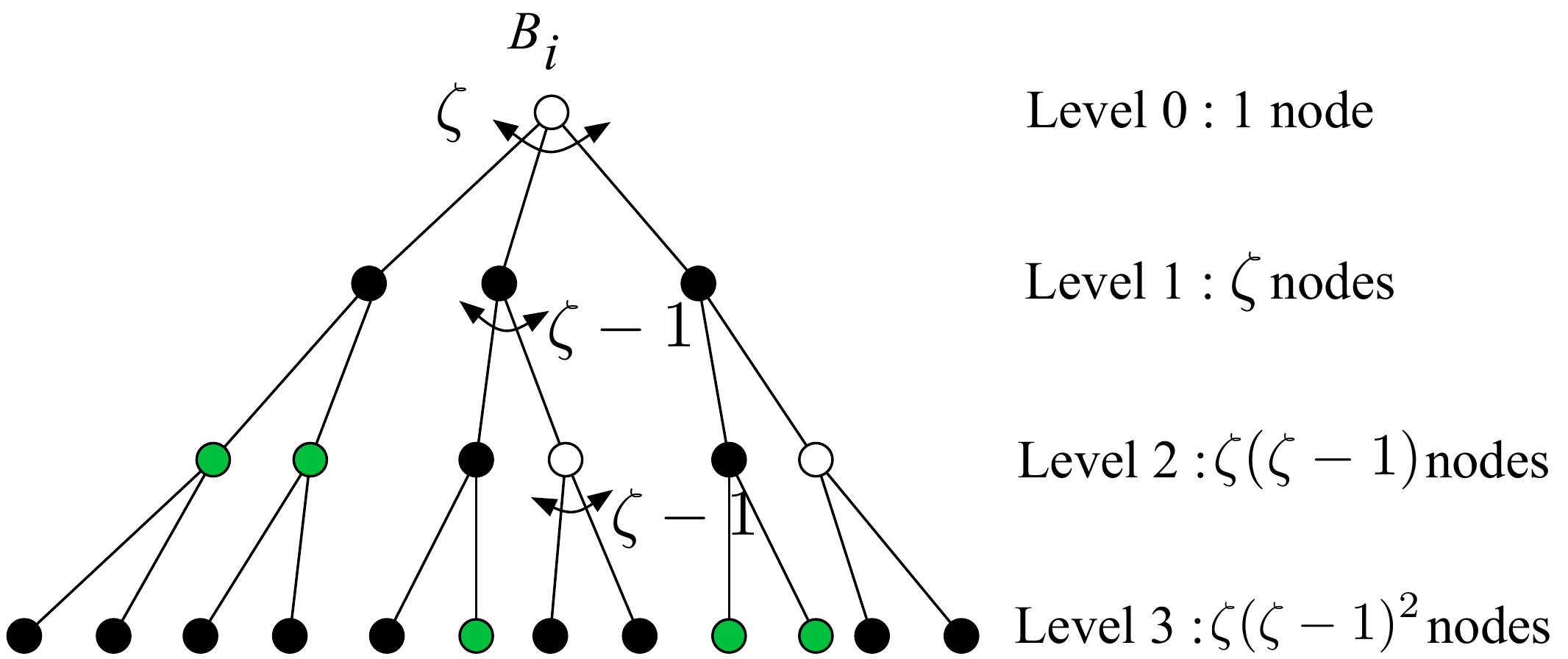}
\caption{An example decoding neighborhood of a message-node (denoted by $B_i$) for $\zeta=3$ (chosen for ease of illustration) after $3$ decoding iterations. The presence of cycles would only decrease the neighborhood size. } 
\label{fig:decodingnbd}
\end{center}
\end{figure}

\begin{definition} \label{def:nbd}
The \textit{neighborhood size} $n_i$ for the $i$-th message bit $B_i$, after $l$ decoding iterations, is the number of channel output nodes that the message node can receive messages from (directly or relayed; see Fig.~\ref{fig:decodingnbd}). The \textit{maximum neighborhood size}, denoted by $n$, is the maximum of the neighborhood sizes over all message nodes.
\end{definition}
\vspace{0.1in}

Two channels are considered in this paper. The first is an AWGN channel with complex Gaussian thermal noise of noise-variance $\sigma_0^2=kT$ per complex-sample, where $k$ is the Boltzmann constant, $T=300$ Kelvin is the room-temperature, and $W_{used}$ is the bandwidth being used. The second is an  AWGN channel where the transmitter uses QPSK symbols and the receiver performs a hard decision on the $I$ and $Q$ channel outputs before decoding. The resulting channel has two parallel Binary Symmetric Channels (BSCs) of crossover probability governed by the transmit power relative to the thermal noise.

%


\subsection{Lower bounds on the number of decoding iterations and decoding power}
\label{sec:lowbound}
In this section, we provide lower bounds on the number of iterations and the required decoding power in the VLSI model of decoding for decoding any code given the rate $R_{ch}$ and the desired error probability $\pe$. These bounds reveal that the decoding neighborhoods must grow unboundedly as the system tries to approach capacity. Since the size of decoding neighborhoods is directly related to the number of iterations, these bounds also yield bounds on the number of iterations, and hence also on the decoding power consumption and total power consumption. The total power consumption bounds are then optimized numerically to obtain plots of the optimizing transmit power and the total power as the average probability of bit-error goes to zero. Our bounds predict that while the optimizing transmit power can stay bounded in this limit, the decoding power must diverge to infinity. 

\subsubsection{Lower bounds on the probability of error as a function of maximum neighborhood size} \label{sec:basicbounds}

This section provides lower bounds on $\pe$ error probability as a function of the maximum neighborhood size $n$. These bounds build on the ``sphere-packing'' analysis for error-probabilities of block-codes as a function of their blocklength~\cite{BlahutHypothesis}. In message-passing decoding, the visible universe for a message-node is not the entire block, but just the decoding neighborhood (see Fig.~\ref{fig:decodingnbd}). 


Because of space-limitations, rigorous derivations of these bounds appear in a companion paper~\cite{ComplexityITPaper}. The intuition behind these bounds is as follows. Analogous to ``sphere-packing'' analysis for blocklength~\cite{BlahutHypothesis}, we first show that the average probability of error for any code must be significant if the channel behaves atypically in a manner that the \textit{effective} capacity of the resulting atypical channel falls below the target rate. We then observe that for a bit to be decoded in error, such atypical behavior is not required for the entire block, but just the visible universe, \textit{i.e.} the decoding neighborhood of the bit~\cite{grover}. Since the probability of this atypical behavior falls at best exponentially with the neighborhood size, so does the error probability of the bit. The precise statements of the bounds now follow.
\begin{theorem}[from~\cite{ComplexityITPaper}] 
\label{thm:basicBSCbound}
Consider a BSC with crossover probability $p < \frac{1}{2}$. Let $n$
be the maximum size of the decoding neighborhood of any individual
message bit. The following lower bound holds on the average probability of bit
error. 
\begin{equation}
\label{eq:peip}
\pe \geq  \sup_{C_{bsc}^{-1}(R) < g \leq \frac{1}{2}}\frac{h_b^{-1}(\delta_{bsc}(g))}{2}
2^{-nD\left(g\|p\right)}\left(\frac{p(1-g)}{g(1-p)}\right)^{\epsilon \sqrt{n}}
\end{equation}
where $h_b(\cdot{})$ is the binary entropy
function, $D(g\|p) = g\lo{\frac{g}{p}} + (1-g)\lo{\frac{1-g}{1-p}}$ is
the KL-divergence, and  $\delta_{bsc}(g)= 1- \frac{C_{bsc}(g)}{R_{ch}}$, where $C_{bsc}(g) = 1 - h_b(g)$ and $\epsilon = \sqrt{\frac{1}{K(g)}\lo{\frac{2}{h_b^{-1}(\delta_{bsc}(g))}}} $where $K(g) = \inf_{0<\eta<1-g} \frac{D(g+\eta \|g)}{\eta^2}$.
\end{theorem}
\begin{theorem}[from~\cite{ComplexityITPaper}]
\label{thm:basicAWGNbound}
For the AWGN channel and the decoder model in
Section~\ref{sec:vlsi}, let $n$ be the  
maximum size of the decoding neighborhood of any individual
message bit. The following lower bound holds on the average
probability of bit-error.
\begin{equation} \label{eq:lbnawgn}
\pe \hspace{-0.05in} \geq  \hspace{-0.1in}\sup_{\sigma_G^2:\;C_{awgn}(\sigma_G^2) < R_{ch}}\hspace{-0.3in}
\frac{h_b^{-1}\left(\delta_{awgn}(\sigma_G^2)\right)}{2}
\exp\left(-nD(\sigma_G^2\|\sigma_0^2)
- \sqrt{n} \left(\frac{3}{2} + 2 \lon{\frac{2}{h_b^{-1}(\delta_{awgn}(\sigma_G^2))}}\right) 
\left(\frac{\sigma_G^2}{\sigma_0^2}-1\right)\right)
\end{equation}
where $\delta_{awgn}(\sigma_G^2) = 1-C_{awgn}(\sigma_G^2)/R_{ch}$, the capacity $C_{awgn}(\sigma_G^2)=\frac{1}{2}\lo{1+\frac{\rxpower}{\sigma_G^2}}$, and the KL divergence $D(\sigma_G^2\|\sigma_0^2) = \frac{1}{2}\left[\frac{\sigma_{G}^2}{\sigma_0^2}-1-\ln\left(\frac{\sigma_{G}^2}{\sigma_0^2}\right) \right]$. 
\end{theorem}
Observe that the right-hand sides of~\eqref{eq:peip} and~\eqref{eq:lbnawgn} are monotonically decreasing in the maximum neighborhood size $n$. For a specified bit-error probability $\pe$, the equations can thus be solved numerically to obtain lower bounds on $n$. These bounds are then used to obtain lower bounds on the number of iterations. Numerical evaluations are in Section~\ref{sec:joint}.

We can get a sense of the qualitative behavior of these bounds by considering the limit $\pe\rightarrow 0$, for which $n$ must diverge to infinity. Taking $\lon{\cdot}$ on both sides of~\eqref{eq:lbnawgn}, for small $\pe$, the term $nD(\sigma_G^2\|\sigma_0^2)$ dominates the other terms
in the RHS. Further, $\sigma_G^2$ can be taken close to ${\sigma_G^*}^2$ that satisfies $C_{awgn}({\sigma_G^*}^2)=R_{ch}$. The other two terms decrease to zero relatively slowly, yielding~\cite{ComplexityITPaper}
\begin{equation}
\label{eq:approx}
n\gtrsim \frac{\lon{1/\pe}}{D({\sigma_G^*}^2\|\sigma_0^2)}
\end{equation}
for the AWGN channel. Similarly, for the BSC we get,
\begin{equation}
\label{eq:approx2}
n\gtrsim \frac{\lo{1/\pe}}{D(g^*\|p)},\; \text{where}\;C_{bsc}(g^*) = R_{ch}.
\end{equation}
It is well known (see, for instance,~\cite[Problem 5.23]{Gallager}) that the divergence terms $D({\sigma_G^*}^2\|\sigma_0^2)$ and $D(g^*\|p)$ behave like $K(C-R_{ch})^2$ in the limit of low error probability, where $C$ is the capacity of the underlying channel, and $K$ is a constant that can depend on the channel and is closely related\footnote{While our bounds essentially capture the correct channel dispersion term for the BSC, a technical difficulty limits us from doing the same for the AWGN channel: the average power of channel inputs for the PEs in the neighborhood could potentially be very different from the average power for the block. We therefore underestimate the true dispersion.} to the ``channel dispersion''~\cite{PolyanskiyDispersion}. The neighborhood size, therefore, must diverge to infinity as the error probability converges to zero or the rate approaches capacity.



\subsubsection{Joint optimization of the weighted total power}
\label{sec:joint}

Let the number of decoding iterations be denoted by $l$. The number of computational nodes can be lower bounded by $m$, the number of received channel outputs. Since each node consumes $E_{node}$ joules of energy in each iteration, the decoding energy $E_{D}$ is lower bounded by 
\begin{equation}
E_D\geq E_{node}\times m\times l.
\end{equation} 
While sphere-packing tools allow us to investigate the impact of random channel-fluctuations, there is no channel in encoding. Therefore, the sphere-packing based lower bound techniques do not seem to apply directly. Further, empirical evidence suggests that it is significantly smaller than the decoding power~\cite{marcu}. Thus we assume that encoding is ``free''. This results in the following lower bound on the weighted total power
\begin{eqnarray}
\label{eq:totalpower}
P_{total}&\geq& P_T +  \frac{\xi_D E_{node} \times m \times l }{T_{dec}}\\&=&\xi_T \rxpower +  \frac{\xi_D E_{node} \times m \times l }{T_{dec}},
\end{eqnarray}
where $T_{dec}=\frac{k}{R_{dec}}$ is the time consumed in decoding. Thus,
\begin{eqnarray}
\label{eq:totalpower2}
P_{total}&\geq &\xi_T \rxpower +  \frac{\xi_D E_{node}mlR_{dec}}{k}\\
& =&\xi_T \rxpower +  \frac{\xi_D E_{node}lR_{dec}}{R_{ch}}.
\end{eqnarray}
We now need to lower bound the number of iterations $l$. This bound is derived by understanding how fast the visible universe for each bit can increase with the number of iterations. After the first iteration, each node has communicated with at most $\zeta$ other neighbors. In each subsequent iteration, each neighbor communicates with at most $\zeta-1$ new neighbors (see Fig.~\ref{fig:decodingnbd} for an illustration). The actual number of neighbors may be smaller because a) each node may not have $\zeta$ neighbors, b) each PE may not store a distinct channel output, and c) there might be cycles  (and hence repetition of nodes) in the decoding neighborhoods. Thus, for $l\geq 1$ and $\zeta>2$,
\begin{eqnarray}
\label{eq:nandl}
n&\leq &\sum_{j=0}^{l-1}\zeta (\zeta-1)^j + 1= \zeta\left( \frac{(\zeta-1)^l-1}{\zeta-2}   \right) + 1.\nonumber\\
\text{Thus,}\;\;(\zeta-1)^l &\geq & \frac{\zeta-2}{\zeta}(n-1) + 1\\
&\Rightarrow&\;l\geq  \frac{\lo{\frac{\zeta-2}{\zeta}n+\frac{2}{\zeta}}}{\lo{\zeta-1}}.
\end{eqnarray}
Using~\eqref{eq:totalpower2}, and observing that $l\geq 0$, 
\begin{eqnarray}
\label{eq:ptot}
P_{total}&\geq& \xi_T \rxpower + \frac{\xi_D E_{node} R_{dec}}{R_{ch}} \frac{\left( \lo{\frac{\zeta-2}{\zeta}n+\frac{2}{\zeta}}\right)^+}{\lo{\zeta-1}}.
\end{eqnarray}
Alternatively, when $\zeta=2$, $n\leq 2l+1$, and thus $l\geq \frac{n-1}{2}$. For numerical results in this paper, we will assume that $\zeta=4$. 

The actual maximum neighborhood size $n$ depends on the coding/decoding technique and on $\txpower$.  However, it can be lower bounded for any code and for a specified $\txpower$ by plugging the desired $\pe$ and $\txpower$ into Theorems~\ref{thm:basicBSCbound} and \ref{thm:basicAWGNbound}. For a fixed transmit power (and hence a fixed gap from capacity), using~\eqref{eq:approx} the error probability falls at most exponentially in the maximum neighborhood size. Since the neighborhood size can grow at best exponentially in the number of iterations, these bounds show that the error probability can fall at best doubly-exponentially in the number of iterations. Because decoding power scales linearly with the number of iterations under our model of Section~\ref{sec:vlsi}, the decoding power must scale at least doubly-logarithmically with the error-probability.

\subsection{Numerical evaluation}

\begin{figure}[htb]
\begin{center}
\includegraphics[width=3.6in]{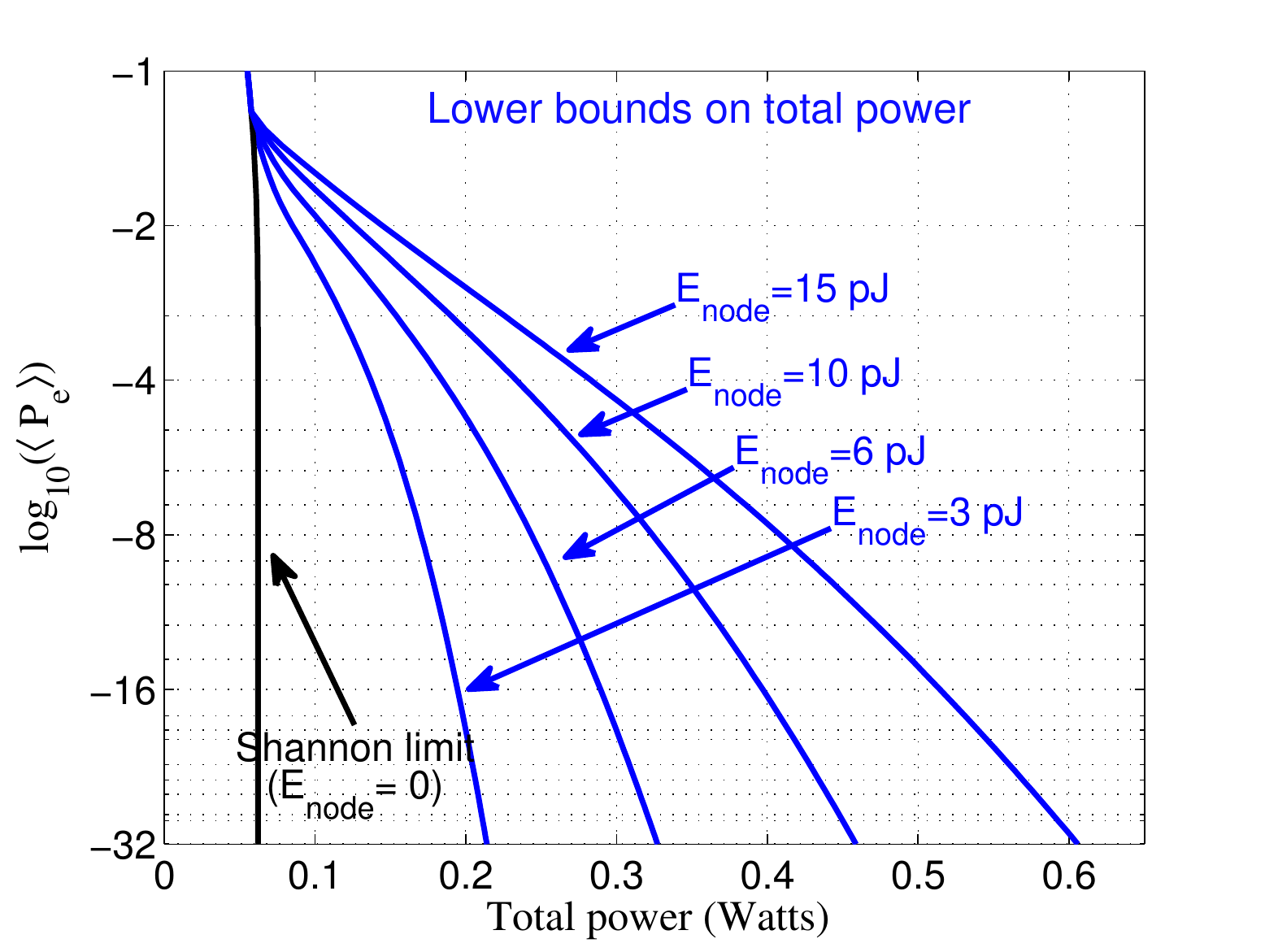}
\caption{The BSC Waterslides: lower bounds on the required total power with QPSK-modulation and hard decisions at the decoder for various values of $E_{node}$. The problem parameters are $r=10$ meters, $R_{data}=1.5$ Gbps, $f_c=60 GHz$, $W=3$ GHz, path-loss exponent $\alpha = 3$. The Shannon limit is a universal lower bound, irrespective of the value of $E_{node}$.}
\label{fig:waterslidebsc}
\end{center}
\end{figure}
\begin{figure}[htb]
\begin{center}
\includegraphics[width=3.6in]{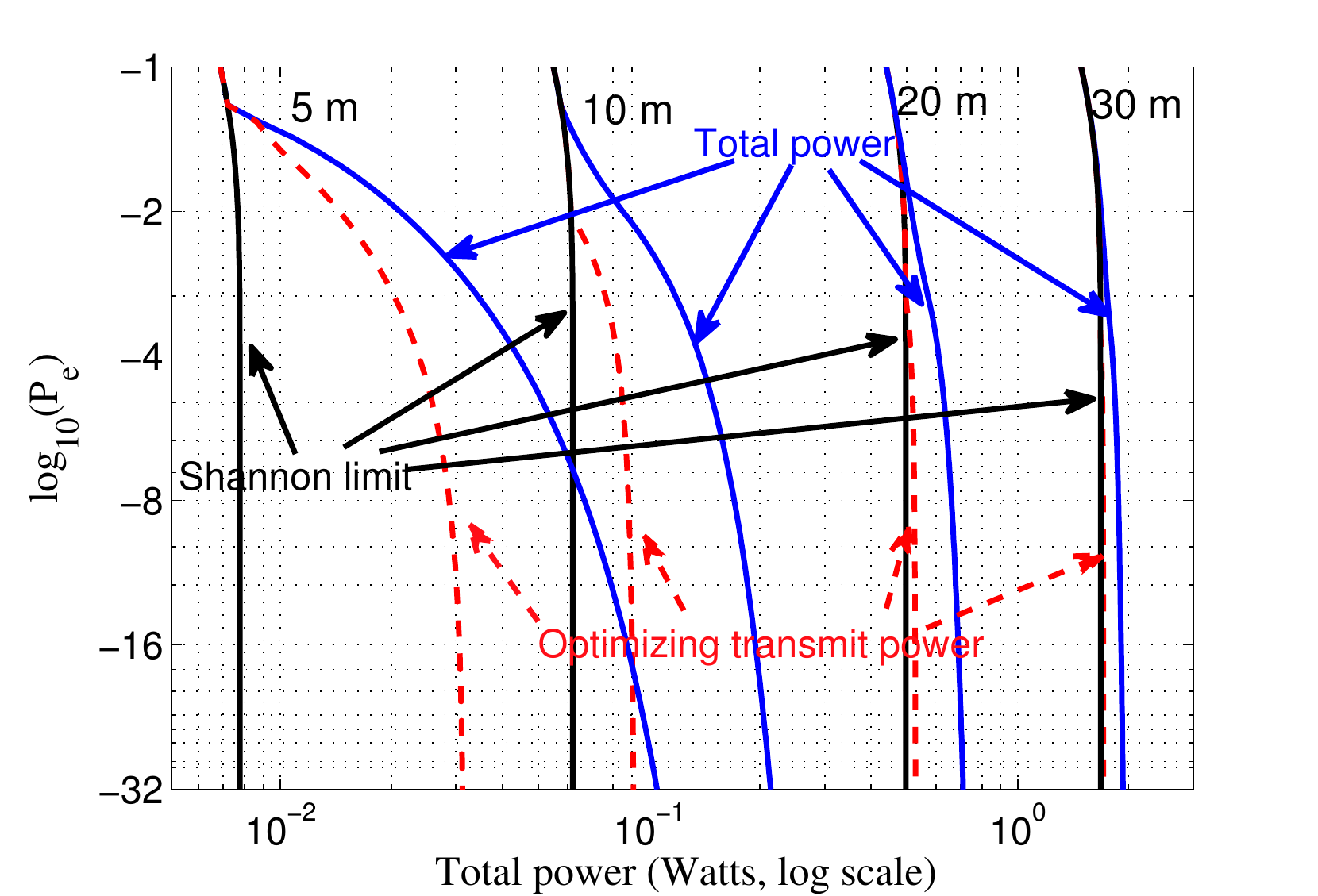}
\caption{The BSC Waterslides: lower bounds on the required total power with QPSK-modulation and hard decisions at the decoder for various values of  the distance $r$. The problem parameters are $E_{node}=3$ picojoules, $R_{data}=1.5$ Gbps, $W=3$ GHz, path-loss exponent $\alpha = 3$. The total power is plotted in log-scale to bring out the relative importance of transmit and decoding power. As expected, the relative importance of decoding power reduces with distance.}
\label{fig:waterslidebscvarious}
\end{center}
\end{figure}

\begin{figure}[htb]
\begin{center}
\includegraphics[width=3.6in]{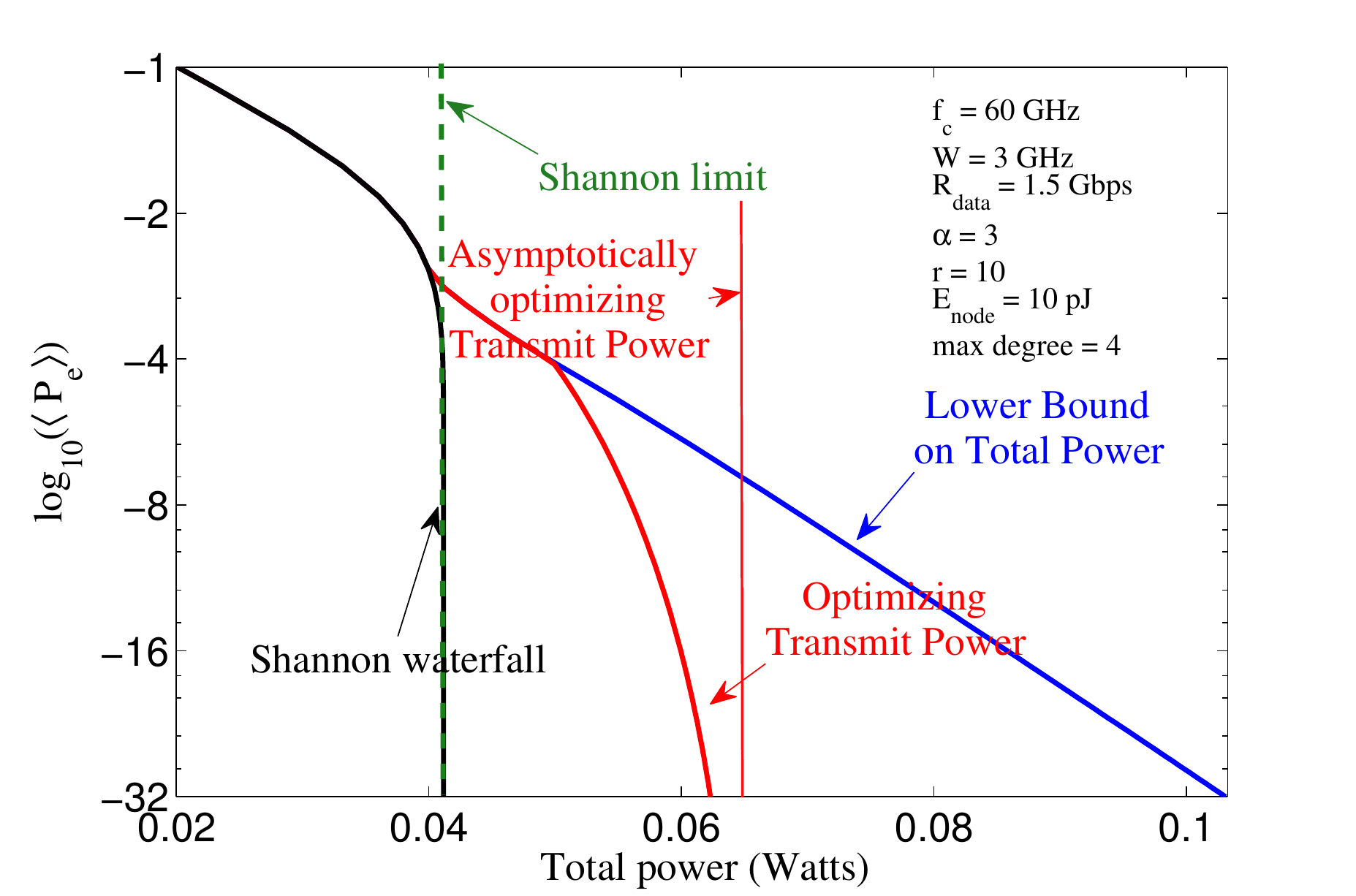}
\caption{The AWGN Waterslide: plots of $\log(\pe)$ (in $\log$ scale to bring out the double-log behavior of decoding and total power) vs lower bounds on required total power for the AWGN channel with the parameters as shown. The initial segment where all the waterslide curves almost coincide illustrates the looseness of the bound since that corresponds to the case of $n = 1$ or when the bound suggests that uncoded transmission could be optimal. However, the bound is too optimistic for uncoded transmission. }
\label{fig:waterslideawgn}
\end{center}
\end{figure}

For numerical evaluation, we assume $f_c=60$ GHz, $W=3$ GHz, the rate $R_{data}=1.5$ Gbps, $\xi_D=1$, path-loss exponent $\alpha=3$, and the maximum connectivity $\zeta=4$. Figures \ref{fig:waterslidebsc}, \ref{fig:waterslidebscvarious} and~\ref{fig:waterslideawgn} show\footnote{Code for all plots in this paper can be found in~\cite{CodeForJSACPaper}.} the total power waterslide curves for fixed technology parameters $\xi_T$, $\zeta$, and $\xi_D$ ($\xi_D$ is assumed to be $1$ in all our curves). The plotted scale is chosen to clearly illustrate the double-exponential relationship between decoding power and probability of error.

From~\eqref{eq:approx} and~\eqref{eq:approx2}, for a given rate $R_{ch}$, if the transmit power $\txpower$ is extremely close to that required for channel capacity to be $R_{ch}$, then the neighborhood size $n$, and so also the number of iterations $l$, would have to be large. From~\eqref{eq:totalpower2}, a large number of iterations require high decoding power. Therefore, the optimized encoder should  transmit at a power larger than that predicted by the Shannon limit in order to decrease the decoding power. It is illustrated in Fig.~\ref{fig:waterslideawgn} and Fig.~\ref{fig:waterslidebscvarious} that this optimizing transmit power is bounded as $\pe\rightarrow 0$. Thus, from~\eqref{eq:peip}, as $\pe\rightarrow 0$, the required neighborhood size $n \rightarrow\infty$. This implies that for any fixed value of transmit power, the power expended at the decoder (and hence the total power) must diverge to infinity as the probability of error converges to zero.

Why is the optimizing transmit power bounded? To get intuition into this, notice that at low error probabilities, 
\begin{eqnarray*}
P_{total}&\overset{\text{using}~\eqref{eq:ptot}}{\geq} & \min_{P_T} P_T +  \frac{\xi_D E_{node} R_{dec}}{R_{ch}} \frac{ \lo{\frac{\zeta-2}{\zeta}n+\frac{2}{\zeta}}}{\lo{\zeta-1}}\\
&\approx &  \min_{P_T} P_T + \gamma \left(    \frac{\lo{\frac{\zeta-2}{\zeta}}}{\lo{\zeta-1}} + \lo{n}  \right)\\
&\overset{\text{using~\eqref{eq:approx} and~\eqref{eq:approx2}}}{ \gtrsim}  & \gamma   \frac{\lo{\frac{\zeta-2}{\zeta}}}{\lo{\zeta-1}} + \min_{P_T}\left\{ P_T  + \gamma \lo{\frac{\lo{1/\pe} }{  K(C(P_T)-R)^2} }\right\}\\
&=&  \gamma   \frac{\lo{\frac{\zeta-2}{\zeta}}}{\lo{\zeta-1}} + \gamma \lo{\frac{\lo{1/\pe} }{K}}+ \min_{P_T}\left\{ P_T   -\frac{2\gamma}{\lon{2}} \lon{C(P_T)-R} \right\}.
\end{eqnarray*}
where $\gamma = \frac{\xi_D E_{node} R_{dec}}{R_{ch}}$. Clearly, any minimizing $P_T$ must satisfy 
\begin{equation}
\label{eq:must}
1 = 2\frac{\gamma}{\lon{2}}\frac{\partial C(P_T)/\partial P_T}{C(P_T)-R}.
\end{equation}
Thus the asymptotically optimizing $P_T$ does not depend\footnote{Does the optimizing $P_T$ depend on the communication range $x$? Even though $x$ does not appear explicitly in~\eqref{eq:must}, it is implicit in the expression $C(P_T)$ through $P_R$.} on $\pe$. Further, it can be shown that in~\eqref{eq:must}, the solution $P_T$ is unique for both AWGN and BSC. 

It is important to note that only the weighted total power curve is a true bound on what a real system could achieve. The constituent $\txpower$ curve is merely an indicator of what the qualitative behavior would be if the true tradeoff behaved like the lower bound. However, our lower bound shows that the decoding power must blow up if you actually approach capacity. 

\vspace{-0.05in}

\subsection{Regular LDPCs attain within a constant factor of the optimal power}
\label{sec:ldpc}

\vspace{-0.1in}

\begin{figure}[htb]
\begin{center}
\includegraphics[width=3.6in]{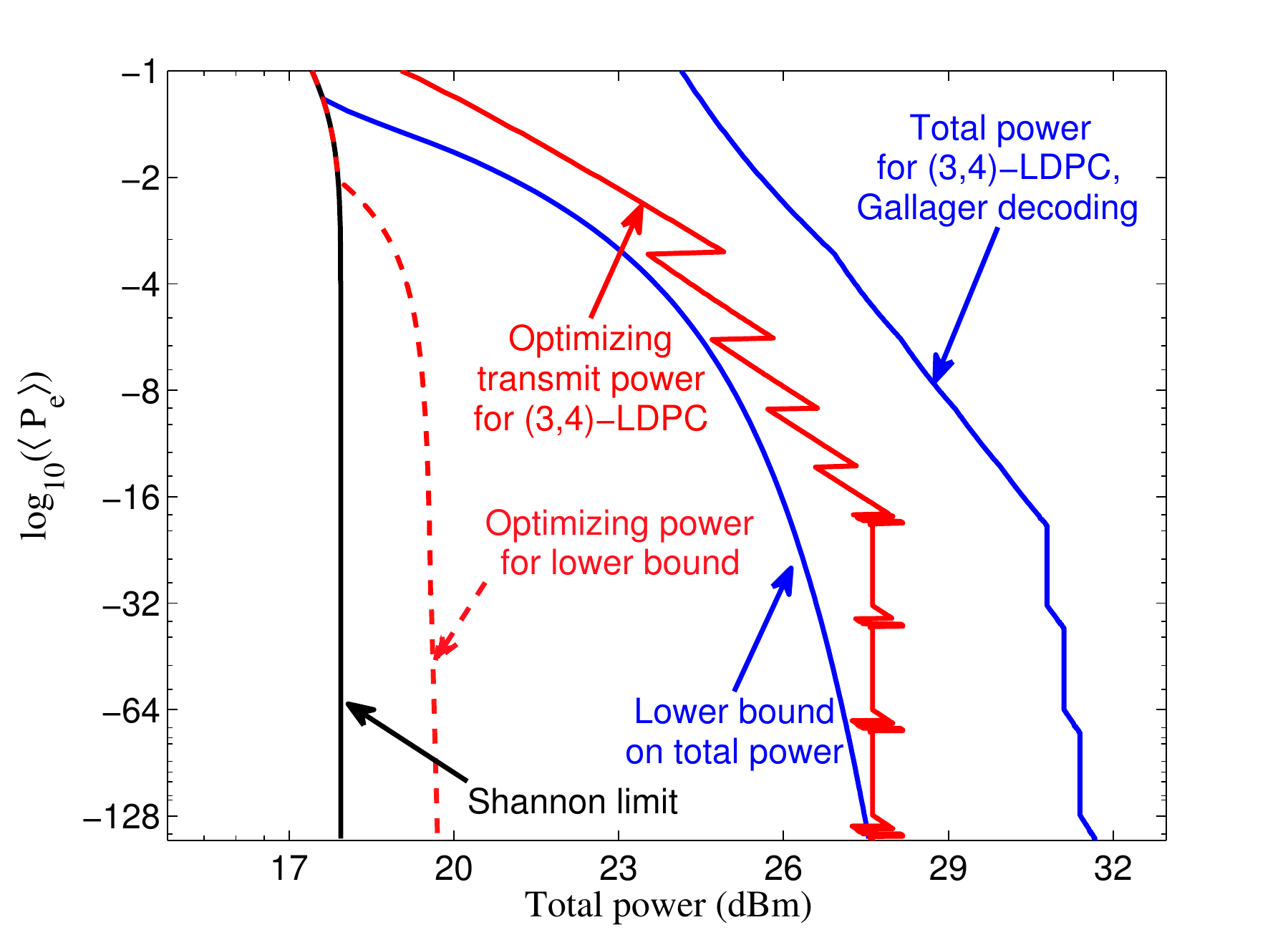}
\caption{The LDPC waterslides: the figure shows the waterslide curve for a $(3,4)$-regular LDPC code (rate $1/4$ bits per channel-use) decoded using the Gallager-B algorithm over a BSC~\cite{urbankecapacity} for $E_{node}=10$ pJ. The appropriate lower bound from Fig.~\ref{fig:waterslidebsc} is also plotted (on log-scale for power). The total power has order-optimal behavior in that the difference between the power achieved by the LDPC code and the lower bound is bounded in dB-scale irrespective of $\pe$ (it is to demonstrate this, we plot the total power for unrealistically low error-probabilities). The decoding algorithm is Gallager B~\cite{urbankecapacity} that passes only one-bit messages along each edge, and requires elementary computations (thresholding at the variable nodes, XOR's at the check nodes) at the PEs. Also shown is the optimal transmit power, which has a saw-tooth behavior because of integer effects. Importantly, the optimal transmit power is bounded. Since the code does not operate close to the channel capacity, the required  transmit power is significantly larger than the optimizing power in the lower bound. }
\label{fig:ldpcwaterslide}
\end{center}
\end{figure}

How far do current constructions operate from this lower bound? In Fig.~\ref{fig:ldpcwaterslide}, the total power required to decode a regular (3,4)-LDPC code using a simple $1$-bit message-passing decoding algorithm called Gallager-B decoding (which is the same as Gallager-A for a (3,4)-LDPC~\cite{urbankecapacity}) is plotted\footnote{We note that the curves are derived assuming that there are no error-floors, that is, the blocklength is infinite and the codes are ``random'' LDPCs designed using the socket-construction of~\cite{urbankecapacity}.} along with the lower bound. Interestingly, this upper bound is separated from the lower bound by a bounded number of dBs even as the error probability falls to zero, suggesting that this (3,4)-LDPC code might be order optimal. Further, as predicted, the optimal transmit power is bounded, but it exhibits a saw-tooth behavior because the number of iterations is an integer for an actual decoder. 

How general is this order-optimality? Our lower bounds on decoding power suggest that the decoding power scales approximately as $\log\frac{\log\frac{1}{\pe}}{(C-R)^2}$. If we are operating at a finite gap from capacity (as any practical code does), then in order to have order optimal decoding power (as $\pe\rightarrow 0$), our code must have decoding power that scales as $\log\log\frac{1}{\pe}$. It is well known~\cite{Lentmaier05} that for regular LDPCs, and indeed for any randomized LDPC code with no degree-2 variable nodes, the error probability falls doubly exponentially with the number of iterations, \textit{i.e.} $l=\Theta\left(\log{\log{\frac{1}{\pe}}}\right)$ in the asymptotic limit of infinite blocklength. Thus regular LDPCs have order optimal \textit{decoding} power. 

What about \textit{total} power? Although regular LDPCs codes do not achieve capacity under belief-propagation-based message-passing decoding~\cite{burshteinBP}, they can be used to communicate reliably at non-zero-rate~\cite{Lentmaier05}, that is, there exists a \textit{finite} transmit power $P_T$ for which $\pe\rightarrow 0$ as $l\rightarrow\infty$  (in the limit of infinite blocklengths), as long as the variable node degree is greater than $2$~\cite{Lentmaier05}. Thus, we only require a constant increase in transmit power. This constant depends on the code's gap from capacity, but is bounded even as $\pe\rightarrow 0$. 

Thus, as $\pe\rightarrow 0$, regular LDPCs indeed require order-optimal total power (within bounded dBs of the optimal). 
However, in the case of (3,4)-LDPC, the gap between the upper and lower bounds is still about $4.8$ dB. We emphasize that this difference is not because of the required increase in transmit power: that increase is only additive and its effect on total power will die down to zero (in dB sense) as $\pe\rightarrow 0$. Partly, there is a gap because we do not count the power consumed by check nodes in the lower bound. This accounts for a looseness of about 2.43 dB. The number of output nodes a (3,4)-LDPC decoder reaches in two clock-cycles is $2\times 3 = 6$. The lower bound assumes that the number of output nodes reached in two clock-cycles is $3\times 3 + 3=12$. This leads to a further loss of $\frac{\lo{12}}{\lo{6}}\approx 1.42$ dB. We suspect that the remaining $0.95$ dB loss is likely due to the use of the Gallager-B algorithm, rather than full belief-propagation, for decoding the LDPC code. 

\vspace{-0.15in}

\section{A collection of point-to-point links}
\label{sec:collection}

\vspace{-0.1in}
\subsection{System model for a collection of point-to-point links}
\label{sec:collectionmodel}

\begin{figure}[htbp] 
   \centering
   \includegraphics[width=2.25in]{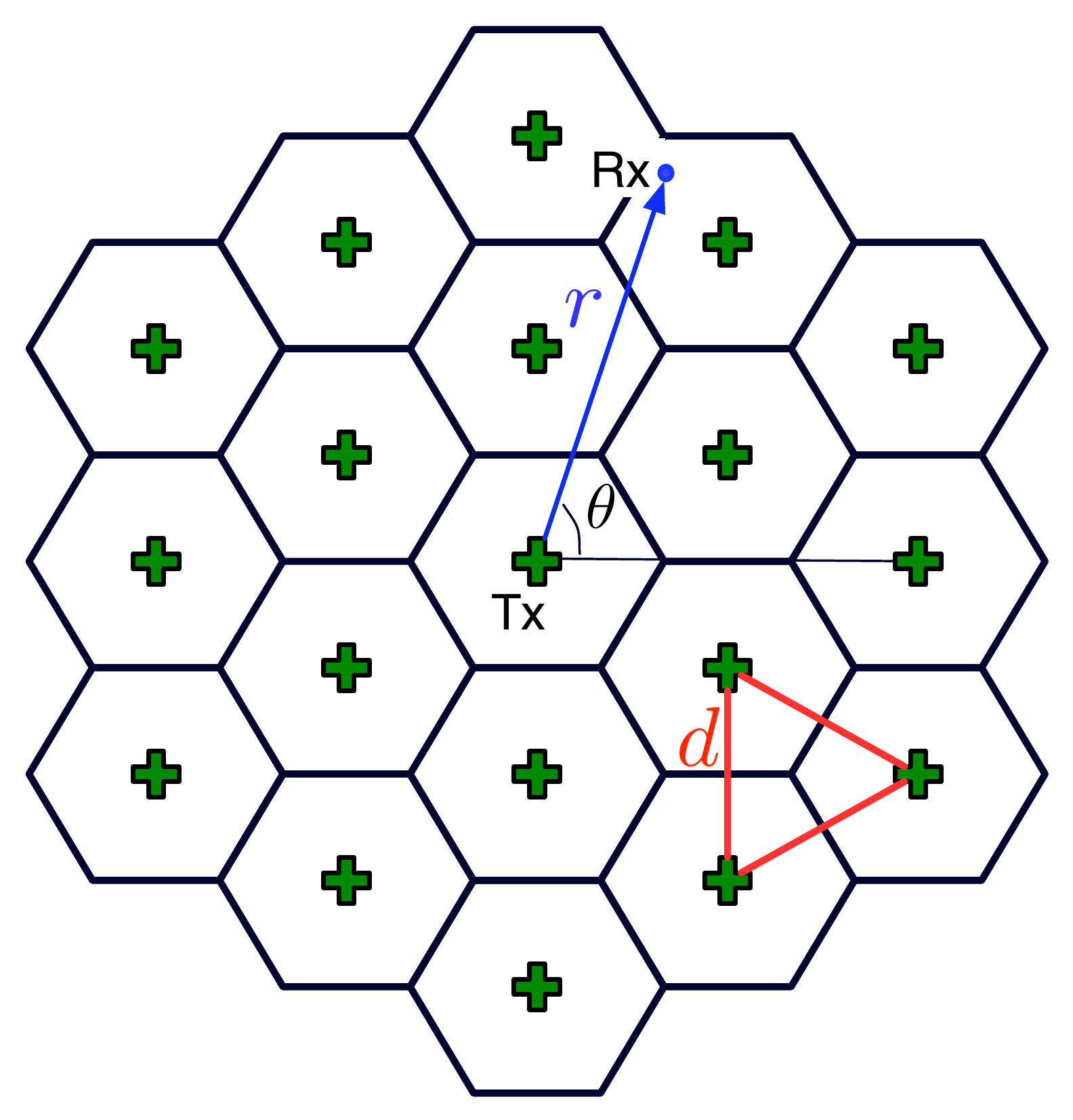} 
   \caption{We consider spatial networks where the transmitters lie on a triangular grid (efficient packing). The density of the transmitters is calculated as follows: the area of an equilateral triangle is $\frac{\sqrt{3}}{4}d^2$, and each triangle contains a total of half-a-transmitter (3 transmitters on each vertex, each shared with 6 other triangles). This gives a density of $\frac{2}{\sqrt{3}d^2}$. }
   \label{fig:gridLattice}
\end{figure}

In this section we consider a situation where the system is assumed to be a collection of point-to-point links. We further assume that the links do not cooperate, and they treat the collective interference from other links as Gaussian noise. If the transmitters are modeled as placed randomly and uniformly, they could be arbitrarily close to each other. These situations are avoided in practice by using a MAC protocol to have some minimum separation between active transmitters~\cite{BaccelliMonograph}. This pushing away of neighbors reduces the interference, thereby allowing for communication at higher rates. The resulting topology is often modeled using a Mat\'{e}rn hard-core process~\cite{HaenggiSurvey}, or (for analytical simplicity) using a regular grid model (for instance, a square-grid or a triangular-grid~\cite{HaenggiNOW}). In this paper, we assume that the transmitters lie on a triangular grid, as shown in Fig.~\ref{fig:gridLattice}. Nearest transmitters are separated by a distance $d$. 

Each node transmits at the same power $\txpower$ to its receiver located at a distance $r$ from the transmitter. This distance is assumed to be fixed, and does not scale with the density of these transmitter-receiver pairs. The communication rate $R_{data}$ (bits per second) and the desired bit-error probability are assumed to be fixed, and equal for all links.

The question we are interested in is: given a particular total power, what strategies allow us to support the maximum number of communication links? In particular, to what extent does the core insight from the point-to-point problem --- that the code should operate at a gap to capacity --- still hold? 

To address the problem in the simplest possible setting, we consider the case of multiple transmitters sending messages to their respective receivers in which :
\begin{itemize}
\item no multi-hop relaying is allowed (unlike that in~\cite{GuptaKumar}). 
\item there is no use of cooperative interference-management strategies (such as those in~\cite{CadambeJafar}) beyond frequency-reuse.
\item the aggregate interference is assumed to behave like additive white Gaussian noise. 
\item the rate of each link is assumed to be fixed at $R_{data}$.
\end{itemize}

\begin{figure}[htbp] 
   \centering
   \includegraphics[width=3.6in]{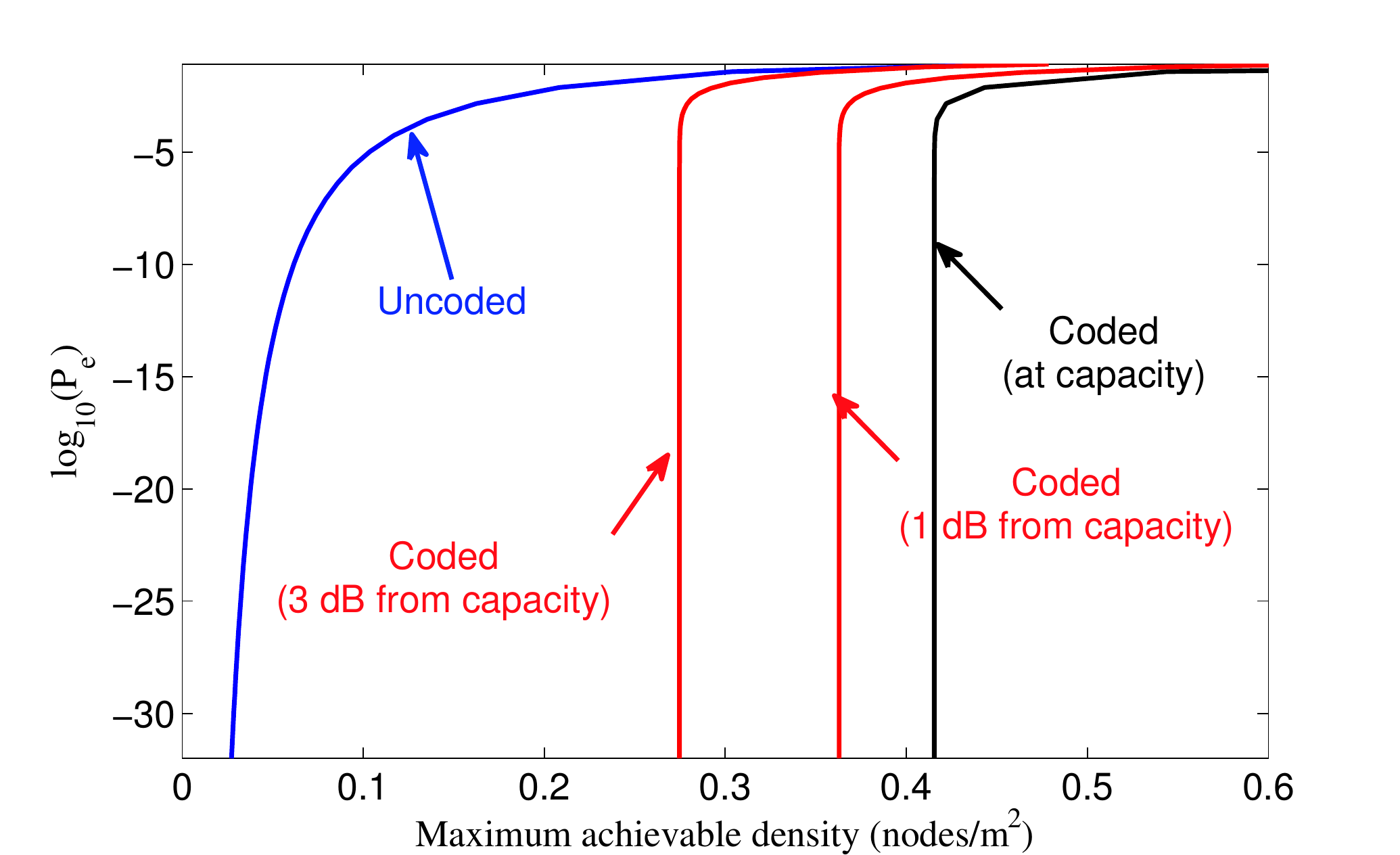} 
   \caption{A plot of achievable densities with decreasing bit-error probability for a rate of $R_{data}=1.5$ Gbps, path-loss exponent $\alpha=3$, bandwidth $W=3$ GHz, central frequency $f_c= 60$ GHz, and distance $r=1$ meter between the transmitters and their receivers, and angle $\theta=0$ (see Fig.~\ref{fig:gridLattice}). The plot shows the maximum attainable density with \textit{arbitrarily large} (but equal) transmission powers. The Shannon-waterfall is reflected as another waterfall  for coded transmissions, yielding a non-zero transmitter-receiver-pair density even as the desired error probability decreases to zero. Similar behavior is demonstrated by codes that operate a few dBs away from capacity. In contrast, because the transmit power for uncoded transmissions must diverge to infinity as $\pe\rightarrow 0$, the density with uncoded transmissions decreases to zero. }
   \label{fig:reversewaterfall}
\end{figure}

For this setup, more meaningful than the waterfall curve in Fig.~\ref{fig:waterfall1} is Fig.~\ref{fig:reversewaterfall}, that plots the maximum \textit{density} of transmitter-receiver pairs that can be supported for a given error probability (the generation of these plots is explained in Section~\ref{sec:largepower}).  The waterfall in Fig.~\ref{fig:waterfall1} translates into a non-zero density of simultaneously active transmitter-receiver pairs whose simultaneous operation can be supported using coded transmissions, even in the limit of tiny bit-error probabilities. By contrast, under the same limit, the supportable density using uncoded transmissions decreases to zero. Because the tolerated interference in uncoded transmissions must go to zero as $\pe\rightarrow 0$, the transmitters are forced to be far from each other. The maximum density plot of Fig.~\ref{fig:reversewaterfall} is obtained in the limit of infinite power. Since decoding power, while substantial, causes no interference\footnote{In the $60$-GHz band, because the operating wavelength is small (less than a centimeter), global interconnects in a decoding chip could potentially act as radiating antennas. Although the impact of the resulting interference may be worthy of consideration, it is ignored in this paper. }, it has no impact on this maximum density.

Because the density of transmitter-receiver pairs is of primary interest, we allow for the transmitters to be stacked closely together so that  potentially interfering transmitters can exist in the middle of a transmitter-receiver pair. Consider a smart phone wirelessly tethered to a farther laptop while a nearer bluetooth headset is communicating to its own laptop. 
The thermal noise observed by a receiver operating in bandwidth $W_{used}$ has a power spectral density of intensity $\frac{kT}{2}$ in each dimension. A given distance $d$ between transmitters immediately translates into the following densities (as explained in Fig.~\ref{fig:gridLattice})

\vspace{-0.35in}

\begin{equation}
\rho_{tri}=\frac{2}{\sqrt{3}d^2}.
\end{equation}
Let $x(i) = id-r\cos\theta$, and $y(j)=j\frac{\sqrt{3}}{2}d - r\sin\theta $ for integers $i$ and $j$, where $\theta$ is as shown in Fig.~\ref{fig:gridLattice}. Then the total interference is given by  
\begin{eqnarray}
\label{eq:interferenceTri}
\nonumber I(\txpower,\lambda,r,d)=  \hspace{-0.2in}\underset{\scriptsize  \begin{array}c i,j=-\infty,\\(i,j)\neq (0,0),\;j\;\text{even}\end{array}}{\sum^\infty } \hspace{-0.2in} P_R\left(\txpower, \sqrt{x(i)^2 + y(j)^2}  \right)+ \hspace{-0.2in} \underset{\scriptsize \begin{array}c i,j=-\infty,\\(i,j)\neq (0,0),\;j\;\text{odd}\end{array}}{\sum^\infty }  \hspace{-0.2in} P_R\left(\txpower, \sqrt{\left(\frac{d}{2}+x(i) \right)^2 + y(j)^2 } \right).
\end{eqnarray}
Interference terms do not have closed-form expressions here, unlike those in~\cite{HaenggiNOW}, because we consider non-zero distances between a transmitter and its receiver which leads to asymmetric terms in the summation. 

Following the work of Alouini and Goldsmith~\cite{AlouiniGoldsmith}, we allow both coded and uncoded transmissions to split the band into multiple sub-bands (of equal bandwidth, each allocated to a different user) in order to reduce co-channel interference while keeping density high. The multiple bands are noninteracting worlds that are assumed to have the same grid structure. The distance $d$ is redefined to be the distance between the nearest transmitters transmitting \textit{in the same band}.


\vspace{-0.2in}

\subsection*{Attained density for uncoded transmission}

\vspace{-0.1in}

For uncoded transmission, we assume that the transmission uses quadrature phase-shift keying (QPSK) modulation with Gray encoding. The bandwidth occupied by the transmissions is assumed to be equal to the data-rate (assuming ideal sinc pulse-shapes). 

If the total available bandwidth is larger than the data rate, the users will obviously split the entire band amongst themselves to reduce interference. The number of sub-bands therefore equals $B=W/R_{data}$. Allowing for this frequency reuse, we obtain the following expression for bit-error probability
\begin{equation}
\pe = \mathbb{Q}\left(  \sqrt{\frac{2P_R(\txpower,r)}{ \frac{kTW}{B}  + I(\txpower,\lambda,r,d)   }}  \right),
\end{equation}
where $I(\txpower,\lambda,r,d)$ is the interference function given by~\eqref{eq:interferenceTri}. For a fixed $\pe$, one can now calculate the required distance $d$ and the maximum density $\rho$ of transmitters that will support rate $R_{data}$.


\vspace{-0.3in}

\subsection*{Attained density for coded transmissions}

\vspace{-0.1in}

We assume that the entire band is split into $B$ equal-sized sub-bands. The maximum allowed interference is now given by the inequality
\begin{equation}
\label{eq:wb}
\frac{W}{B} \lo{1+{\frac{P_R(\txpower,r)}{ \frac{kTW}{B}  + I(\txpower,\lambda,r,d)   }}} \geq R_{data} (1-h_b(\pe)),
\end{equation}
where $I(\txpower,\lambda,r,d)$ is the interference function given by~\eqref{eq:interferenceTri}. The term $1-h_b(\pe)$ in the right-hand side of~\eqref{eq:wb} is to account for the fact that at finite \textit{bit}-error probabilities, one could conceivably communicate at rates above capacity\footnote{Although this bound is present implicitly in the other lower bounds in this paper, it appears explicitly only here. The proof is standard, and can be found in~\cite{ComplexityITPaper}.}. 
For fixed $\pe$ and $R_{data}$, the distance $d$ and density $\rho$ can again be calculated. Notice that because of the choice of modulation scheme, there is freedom in the choice of $B$. This freedom is much more curtailed for uncoded transmission, where the bandwidth and rate are intimately tied. There can, however, be flexibility in choice of the constellation which we ignore for simplicity.

\vspace{-0in}

\subsection{Maximum attainable density at infinite power}
\label{sec:largepower}

\vspace{-0.1in}

To understand the limits of what is possible with coding, we first find the asymptotic density in the limit of infinite power. Because decoding power does not pollute, it is ignored in the analysis. In Fig.~\ref{fig:reversewaterfall}, we compare the maximum achievable density of transmitter-receiver pairs  using coded and uncoded transmissions for a fixed rate. Reflecting the waterfall curve, the  attainable density does not decrease to zero for coded transmission even as the error probability decreases to zero, unlike the behavior for uncoded transmissions. Therefore, to support higher densities of high-quality links, the designer must use coded transmissions even if that means incurring a large decoding power cost.

\vspace{-0in}

\subsection{Attainable density at finite total power}
\label{sec:finite}
\vspace{-0.1in}

In practice, the available total power per link is finite. Coding thus needs to be penalized for using decoding power. In particular, at the low densities that are achievable using uncoded transmissions, there is a possibility that uncoded transmissions could use less total power despite needing more transmit power. Further, we must ask whether at densities that require coding, should we use capacity achieving codes?

\begin{figure}[htbp] 
   \centering
   \includegraphics[width=3.6in]{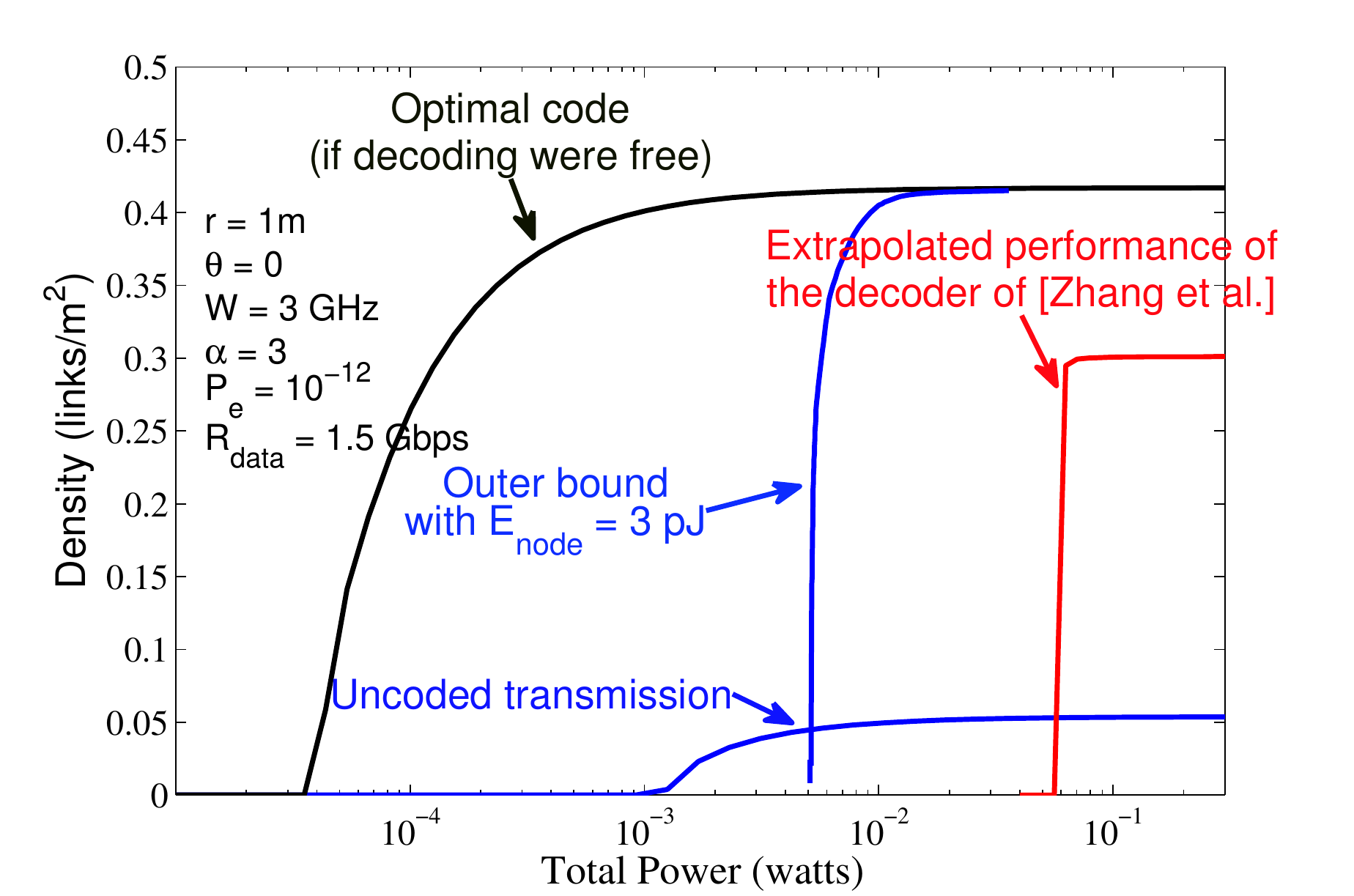} 
   \caption{Comparison of achievable transmitter-receiver pair densities versus transmit power for a rate of $1.5$ Gbps over a triangular network with other parameters as before. Because the required SINR in the code-decoder of~\cite{zhengyaJournal} is rather high ($5.5$ dB for a rate of 0.8125 bits/channel use), there is a substantial gap from the optimal even in the limit of infinite power. Also plotted is an  upper bound on the density attained by an optimal code based on our results in Section~\ref{sec:isolated}, assuming $E_{node}=3$ pJ, which is the approximate value of $E_{node}$ in~\cite[Table V]{zhengyaJournal}. This upper bound also assumes that at least one decoding iteration is performed. The plot shows that for extremely low total power, uncoded transmission is the only feasible strategy if $E_{node}$ cannot be lowered.}
   \label{fig:comparison}
\end{figure}
\begin{figure}[htbp] 
   \centering
   \includegraphics[width=3.3in]{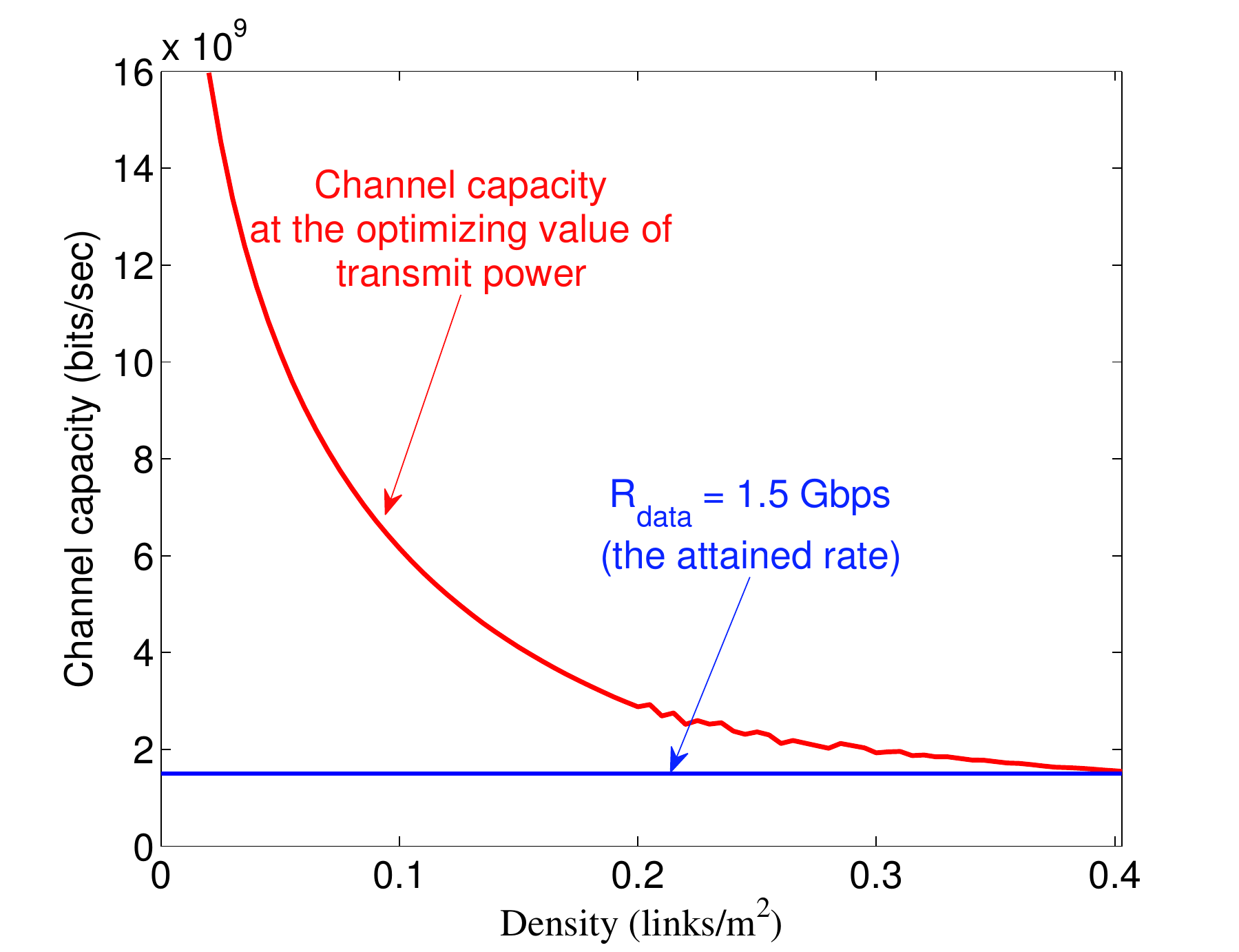}
   \caption{A plot of capacity corresponding to optimizing SINR in the optimal code performance bound using complexity lower bounds in Fig.~\ref{fig:comparison}. A finite gap from capacity is required in order to decrease the decoding power.}
   \label{fig:densityvsgap}
\end{figure}

We plot the performance of the code/decoder pair of~\cite{zhengyaJournal} in Fig.~\ref{fig:comparison}. At low densities, uncoded transmission indeed outperforms coded --- after all, the decoders run at least one iteration, so they require a minimum power to run! As expected, the high densities are only supportable by coded transmission, even though Fig.~\ref{fig:densityvsgap} shows that the codes must still operate at a gap from capacity for any finite power. 

How much could we gain by changing the code? We could certainly improve the maximum attainable density by building codes that approach capacity. The challenge is that decoding power depends on both the code and the decoding architecture. It is here that the lower bounds on power consumption of Section~\ref{sec:isolated} prove useful. These bounds are turned into upper bounds on density for given total power, and are also plotted in Fig.~\ref{fig:comparison}. They show that at low transmit power, uncoded transmission outperforms any code that is decoded with the architecture of~\cite{zhengyaJournal} (and hence has $E_{node}=3$ picojoules~\cite[Table V]{zhengyaJournal}). A more important observation, illustrated in Fig.~\ref{fig:densityvsgap} is that we again should not operate the codes at capacity to optimize the link density given a total power constraint.

\vspace{-0.2in}

\section{Discussions and conclusions}
\label{sec:conclusions}

\vspace{-0.1in}

In this paper, we used a very simple model to account for decoding implementation and decoding power. But even this simplistic model suffices to show that operating close to capacity will fundamentally require a large decoding power. For obtaining deeper insights into design of codes, a more refined modeling of the decoding implementation is required. Implementation models and results for specific code/decoder families (for example, see~\cite{sason}) is needed to complement our fundamental analysis.

In case of an isolated point-to-point link, if the communication distance is small, keeping a  sufficient gap from capacity becomes significantly more important because the decoding power and transmit power are comparable. Nevertheless, the total (transmit+decoding) power must diverge to infinity as $\pe\rightarrow 0$. 

Because we assume almost nothing about the code structure, the bounds here are much more optimistic than those in~\cite{sason} (because of space constraints here, a comparison appears in~\cite{ComplexityITPaper}). However, it is unclear to what extent the optimism of our bound is an artifact of our derivation technique. After all,~\cite{Lentmaier05} does get double-exponential reductions in probability of error with additional iterations, but for a family of codes that does not seem to approach capacity. It is here that code constructions of~\cite{LentmaierITAPaper} may prove useful --- these codes have a doubly-exponential fall in error probability with iterations, while seemingly attaining rates close to capacity.


In an environment where a collection of links is operating simultaneously, the challenge is pictorially captured in Fig.~\ref{fig:comparison}. While improvements in codes to make them capacity approaching will bring the high-power performance of links closer to optimal, low-complexity designs  may outperform uncoded transmission at lower power. As we show, there are tradeoffs between the two, and obtaining improved bounds on this tradeoff is a challenge for information and coding theorists.

\vspace{-0.2in}

\bibliographystyle{IEEEtran}

\bibliography{IEEEabrv,MyMainBibliography,MyMainBibliography2}

\begin{thebibliography}{10}
\providecommand{\url}[1]{#1}
\csname url@samestyle\endcsname
\providecommand{\newblock}{\relax}
\providecommand{\bibinfo}[2]{#2}
\providecommand{\BIBentrySTDinterwordspacing}{\spaceskip=0pt\relax}
\providecommand{\BIBentryALTinterwordstretchfactor}{4}
\providecommand{\BIBentryALTinterwordspacing}{\spaceskip=\fontdimen2\font plus
\BIBentryALTinterwordstretchfactor\fontdimen3\font minus
  \fontdimen4\font\relax}
\providecommand{\BIBforeignlanguage}[2]{{%
\expandafter\ifx\csname l@#1\endcsname\relax
\typeout{** WARNING: IEEEtran.bst: No hyphenation pattern has been}%
\typeout{** loaded for the language `#1'. Using the pattern for}%
\typeout{** the default language instead.}%
\else
\language=\csname l@#1\endcsname
\fi
#2}}
\providecommand{\BIBdecl}{\relax}
\BIBdecl

\bibitem{greencodes}
P.~Grover and A.~Sahai, ``Green codes: Energy-efficient short-range
  communication,'' in \emph{Proceedings of the 2008 {IEEE} Symposium on
  Information Theory}, Toronto, Canada, Jul. 2008.

\bibitem{ISTC10Paper}
P.~Grover, K.~Woyach, H.~Palaiyanur, and A.~Sahai, ``{An interference-aware
  perspective on decoding power},'' in \emph{6th International symposium on
  turbo codes and iterative information processing}, Brest, France, Sep. 2010.

\bibitem{Transistor}
\BIBentryALTinterwordspacing
The miracle month: the invention of the first transistor. [Online]. Available:
  \url{http://www.pbs.org/transistor/background1/events/miraclemo.html}
\BIBentrySTDinterwordspacing

\bibitem{ShannonOriginalPaper}
C.~E. Shannon, ``A mathematical theory of communication,'' \emph{Bell Sys.
  Tech. Jour.}, vol.~27, pp. 379--423, 623--656, Jul./Oct. 1948.

\bibitem{Massey92DeepSpace}
J.~Massey, ``Deep-space communications and coding: A marriage made in heaven,''
  in \emph{Advanced Methods for Satellite and Deep Space Communications:
  Lecture Notes in Control and Information Sciences 182}, J.~Hagenauer,
  Ed.\hskip 1em plus 0.5em minus 0.4em\relax New York: Springer, 1992, pp.
  1--17.

\bibitem{MansourShanbhag}
M.~Mansour and N.~Shanbhag, ``{High-throughput LDPC decoders},'' \emph{IEEE
  Tran. Very Large Scale Integration Systems}, vol.~11, pp. 976--996, 2003.

\bibitem{zhengyaJournal}
Z.~Zhang, V.~Anantharam, M.~Wainwright, and B.~Nikolic, ``An efficient
  {10GBASE-T} ethernet {LDPC} decoder design with low error floors,''
  \emph{Solid-State Circuits, IEEE Journal of}, vol.~45, no.~4, pp. 843 --855,
  Apr. 2010.

\bibitem{zhengyaThesis}
Z.~Zhang, ``{Design of LDPC decoders for improved low error rate
  performance},'' Ph.D. dissertation, UC Berkeley, Berkeley, CA, 2009.

\bibitem{hanzo}
L.~Hanzo, \emph{{OFDM and MC-CDMA for broadband multi-user communications,
  WLANs, and broadcasting}}.\hskip 1em plus 0.5em minus 0.4em\relax Wiley-IEEE
  Press, 2003.

\bibitem{marcu}
C.~Marcu, D.~Chowdhury, C.~Thakkar, J.-D. Park, L.-K. Kong, M.~Tabesh, Y.~Wang,
  B.~Afshar, A.~Gupta, A.~Arbabian, S.~Gambini, R.~Zamani, E.~Alon, and
  A.~Niknejad, ``A 90 nm {CMOS} low-power 60 {GHz} transceiver with integrated
  baseband circuitry,'' \emph{Solid-State Circuits, IEEE Journal of}, vol.~44,
  no.~12, pp. 3434 --3447, Dec. 2009.

\bibitem{goldsmithbahai}
{S Cui, AJ Goldsmith and A Bahai}, ``{Energy Constrained Modulation
  Optimization},'' \emph{{IEEE} Trans. Wireless Commun.}, vol.~4, no.~5, pp.
  1--11, 2005.

\bibitem{massaadJournal}
P.~Youssef-Massaad, L.~Zheng, and M.~Medard, ``Bursty transmission and glue
  pouring: on wireless channels with overhead costs,'' \emph{IEEE Transactions
  on Wireless Communications}, vol.~7, no.~12, pp. 5188 --5194, Dec. 2008.

\bibitem{VerduCost}
S.~Verd\'{u}, ``On channel capacity per unit cost,'' \emph{{IEEE} Trans.
  Inform. Theory}, vol.~36, no.~9, pp. 1019--1030, Sep. 1990.

\bibitem{BurstyHajek}
V.~Subramanian and B.~Hajek, ``Broad-band fading channels: signal burstiness
  and capacity,'' \emph{Information Theory, IEEE Transactions on}, vol.~48,
  no.~4, pp. 809 --827, Apr. 2002.

\bibitem{ModernCodingTheory}
T.~Richardson and R.~Urbanke, \emph{Modern Coding Theory}.\hskip 1em plus 0.5em
  minus 0.4em\relax Cambridge University Press, 2007.

\bibitem{HowardSchlegel}
S.~L. Howard, C.~Schlegel, and K.~Iniewski, ``Error control coding in low-power
  wireless sensor networks: when is {ECC} energy-efficient?'' \emph{{EURASIP}
  Journal on Wireless Communications and Networking}, pp. 1--14, 2006.

\bibitem{urbankecapacity}
T.~J. Richardson and R.~L. Urbanke, ``The capacity of low-density parity-check
  codes under message-passing decoding,'' \emph{{IEEE} Trans. Inform. Theory},
  vol.~47, no.~2, pp. 599--618, Feb. 2001.

\bibitem{lengauer}
T.~Lengauer, ``{VLSI} theory,'' \emph{Handbook of theoretical computer science
  (vol. A): algorithms and complexity}, pp. 835--866, 1990.

\bibitem{thompson}
C.~D. Thompson, ``Area-time complexity for {VLSI},'' in \emph{Proceedings of
  the 11th annual ACM symposium on Theory of computing (STOC)}.\hskip 1em plus
  0.5em minus 0.4em\relax New York, NY, USA: ACM, 1979, pp. 81--88.

\bibitem{ComplexityITPaper}
\BIBentryALTinterwordspacing
A.~Sahai and P.~Grover, ``A general lower bound on the decoding complexity of
  message-passing decoding,'' in \emph{In preparation}, 2010. [Online].
  Available:
  \url{http://www.eecs.berkeley.edu/$\sim$pulkit/papers/ComplexityITPaper.pdf}
\BIBentrySTDinterwordspacing

\bibitem{BlahutHypothesis}
R.~Blahut, ``{Hypothesis testing and information theory},'' \emph{Information
  Theory, IEEE Transactions on}, vol.~20, no.~4, pp. 405--417, 2002.

\bibitem{SahaiDelayBlocklength}
A.~Sahai, ``Why block-length and delay behave differently if feedback is
  present,'' \emph{{IEEE} Trans. Inform. Theory}, pp. 1860 -- 1886, May 2008.

\bibitem{sason}
I.~Sason and G.~Wiechman, ``Bounds on the number of iterations for turbo-like
  ensembles over the binary erasure channel,'' \emph{IEEE Trans. Inf. Theor.},
  vol.~55, no.~6, pp. 2602--2617, 2009.

\bibitem{burshteinBP}
D.~Burshtein and G.~Miller, ``Bounds on the performance of belief propagation
  decoding,'' \emph{IEEE Transactions on Information Theory}, vol.~48, no.~1,
  pp. 112 --122, Jan. 2002.

\bibitem{ISMInterference}
N.~Golmie, ``{Interference in the 2.4 GHz ISM band: challenges and
  solutions},'' \emph{Network. for Perv. Computing}, vol. 500, p.~48, 2005.

\bibitem{thompsonthesis}
C.~D. Thompson, ``A complexity theory for {VLSI},'' Ph.D. dissertation,
  Pittsburgh, PA, USA, 1980.

\bibitem{cole88}
R.~Cole and A.~Siegel, ``Optimal {VLSI} circuits for sorting,'' \emph{J. ACM},
  vol.~35, no.~4, pp. 777--809, 1988.

\bibitem{sinhamultiply}
B.~P. Sinha and P.~K. Srimani, ``A new parallel multiplication algorithm and
  its {VLSI} implementation,'' in \emph{Proceedings of the 1988 ACM 16th annual
  conference on computer science}.\hskip 1em plus 0.5em minus 0.4em\relax New
  York, NY, USA: ACM, 1988, pp. 366--372.

\bibitem{kramerboolean}
M.~R. Kramer and J.~{van\ Leeuwen}, ``The {VLSI} complexity of boolean
  functions,'' in \emph{Proceedings of the Symposium "Rekursive Kombinatorik"
  on Logic and Machines: Decision Problems and Complexity}.\hskip 1em plus
  0.5em minus 0.4em\relax London, UK: Springer-Verlag, 1984, pp. 397--407.

\bibitem{BhattUniversal}
S.~N. Bhatt, G.~Bilardi, and G.~Pucci, ``Area-time tradeoffs for universal
  {VLSI} circuits,'' \emph{Theoretical Computer Science}, vol. 408, no. 2-3,
  pp. 143 -- 150, 2008, excursions in Algorithmics: A Collection of Papers in
  Honor of Franco P. Preparata.

\bibitem{Yao}
A.~C.-C. Yao, ``Some complexity questions related to distributive
  computing(preliminary report),'' in \emph{STOC '79: Proceedings of the
  eleventh annual ACM symposium on Theory of computing}.\hskip 1em plus 0.5em
  minus 0.4em\relax New York, NY, USA: ACM, 1979, pp. 209--213.

\bibitem{UllmanBook}
J.~Ullman, \emph{{Computational Aspects of {VLSI} Design}}.\hskip 1em plus
  0.5em minus 0.4em\relax Computer Science Press, 1983.

\bibitem{Yao2}
A.~C.-C. Yao, ``The entropic limitations on {VLSI} computations (extended
  abstract),'' in \emph{STOC '81: Proceedings of the thirteenth annual ACM
  symposium on Theory of computing}.\hskip 1em plus 0.5em minus 0.4em\relax New
  York, NY, USA: ACM, 1981, pp. 308--311.

\bibitem{arikanBP}
E.~Arikan, ``{A performance comparison of polar codes and Reed-Muller codes},''
  \emph{Communications Letters, IEEE}, vol.~12, no.~6, pp. 447--449, 2008.

\bibitem{Vasudevan}
S.~Vasudevan, C.~Zhang, D.~Goeckel, and D.~Towsley, ``Optimal power allocation
  in wireless networks with transmitter-receiver power tradeoffs,''
  \emph{Proceedings of the 25th IEEE International Conference on Computer
  Communications {INFOCOM}}, pp. 1--11, Apr. 2006.

\bibitem{Gallager}
R.~G. Gallager, \emph{Information Theory and Reliable Communication}.\hskip 1em
  plus 0.5em minus 0.4em\relax New York, NY: John Wiley, 1971.

\bibitem{grover}
P.~Grover, ``Bounds on the tradeoff between rate and complexity for
  sparse-graph codes,'' in \emph{2007 {IEEE} Information Theory Workshop
  ({ITW})}, Lake Tahoe, CA, 2007.

\bibitem{PolyanskiyDispersion}
Y.~Polyanskiy, H.~Poor, and S.~Verdu, ``{Dispersion of Gaussian channels},'' in
  \emph{IEEE International Symposium on Information Theory (ISIT)}.\hskip 1em
  plus 0.5em minus 0.4em\relax IEEE, 2009, pp. 2204--2208.

\bibitem{CodeForJSACPaper}
\BIBentryALTinterwordspacing
P.~Grover, ``Code for ``towards a communication-theoretic understanding of
  system-level power consumption''.'' [Online]. Available:
  \url{http://www.eecs.berkeley.edu/$\sim$pulkit/CodeForJSACSubmission.htm}
\BIBentrySTDinterwordspacing

\bibitem{Lentmaier05}
M.~Lentmaier, D.~V. Truhachev, K.~S. Zigangirov, and D.~J. Costello, ``An
  analysis of the block error probability performance of iterative decoding,''
  \emph{{IEEE} Trans. Inform. Theory}, vol.~51, no.~11, pp. 3834--3855, Nov.
  2005.

\bibitem{BaccelliMonograph}
F.~Baccelli and B.~Blaszczyszyn, ``Stochastic geometry and wireless networks:
  Volume {I} theory,'' \emph{Found. Trends Netw.}, vol.~3, no. 3-4, pp.
  249--449, 2009.

\bibitem{HaenggiSurvey}
M.~Haenggi, J.~G. Andrews, F.~Baccelli, O.~Dousse, and M.~Franceschetti,
  ``Stochastic geometry and random graphs for the analysis and design of
  wireless networks,'' \emph{IEEE J.Sel. A. Commun.}, vol.~27, no.~7, pp.
  1029--1046, 2009.

\bibitem{HaenggiNOW}
M.~Haenggi and R.~K. Ganti, \emph{Interference in Large Wireless
  Networks}.\hskip 1em plus 0.5em minus 0.4em\relax Hanover, MA: Foundations
  and Trends in Communication and Information theory, NOW Publishing, 2008,
  vol.~3, no.~8.

\bibitem{GuptaKumar}
P.~Gupta and P.~R. Kumar, ``The capacity of wireless networks,'' \emph{{IEEE}
  Trans. Inform. Theory}, vol.~46, no.~2, pp. 388 --404, Mar. 2000.

\bibitem{CadambeJafar}
V.~Cadambe and S.~A. Jafar, ``Interference alignment and degrees of freedom of
  the {K}-user interference channel,'' \emph{Information Theory, IEEE
  Transactions on}, vol.~54, no.~8, pp. 3425--3441, Aug. 2008.

\bibitem{AlouiniGoldsmith}
{M-S Alouini and AJ Goldsmith}, ``Area spectral efficiency of cellular mobile
  radio systems,'' \emph{IEEE Transactions On Vehicular Technology}, vol.~48,
  no.~4, 1999.

\bibitem{LentmaierITAPaper}
M.~Lentmaier, G.~Fettweis, K.~Zigangirov, and D.~Costello, ``{Approaching
  capacity with asymptotically regular LDPC codes},'' in \emph{Information
  Theory and Applications Workshop, 2009}.\hskip 1em plus 0.5em minus
  0.4em\relax IEEE, 2009, pp. 173--177.

\end{thebibliography}




\end{document}